\documentclass[11pt]{article}

\usepackage[margin=1in]{geometry}
\usepackage{amsmath,amssymb,amsthm}
\usepackage{graphicx}
\usepackage{bm}
\usepackage{bbold}
\usepackage{mathbbol}
\usepackage{booktabs}
\usepackage{multirow}
\usepackage{adjustbox}
\usepackage{float}
\usepackage{lscape}
\usepackage{rotating}
\usepackage{setspace}
\usepackage{ragged2e}
\usepackage{tikz}
\usepackage{xcolor}
\usepackage[caption=false]{subfig}
\usepackage{algorithm}
\usepackage{algorithmicx}
\usepackage{algpseudocode}
\usepackage{natbib}
\usepackage[colorlinks=true,allcolors=blue]{hyperref}

\newtheorem{theorem}{Theorem}[section]

\newtheorem{remark}[theorem]{Remark}
\newtheorem{definition}[theorem]{Definition}

\def\*#1{\mathbf{#1}}
\DeclareSymbolFontAlphabet{\amsmathbb}{AMSb}
\def\+#1{\amsmathbb{#1}}
\def\##1{\mathbb{#1}}

\newcommand{\calT}{\mathcal T}

\newcommand{\bC}{{\boldsymbol C}}

\newcommand{\bV}{{\boldsymbol V}}

\newcommand{\bX}{{\boldsymbol X}}

\newcommand{\bxi}{{\boldsymbol \xi}}

\newcommand{\bDelta}{{\boldsymbol \Delta}}

\newcommand{\bphi}{{\boldsymbol \phi}}

\title{Multivariate Functional Principal Component Analysis for Mixed-Type mHealth Data: An Application to Mood Disorders}

\author{
Debangan Dey$^{1}$\thanks{Corresponding author: debangan@tamu.edu}
\and Rahul Ghosal$^{2}$
\and Kathleen Merikangas$^{3}$
\and Vadim Zipunnikov$^{4}$\\[1ex]
{\small $^{1}$Department of Statistics, Texas A\&M University, College Station, TX, USA}\\
{\small $^{2}$Department of Epidemiology and Biostatistics, University of South Carolina, Columbia, SC, USA}\\
{\small $^{3}$National Institute of Mental Health, Bethesda, MD, USA}\\
{\small $^{4}$Department of Biostatistics, Johns Hopkins Bloomberg School of Public Health, Baltimore, MD, USA}
}

\begin{document}

\maketitle

\begin{abstract}
    Modern mobile health (mHealth) assessment combines self-reported measures of participants’ health experiences with passively collected health behavior data throughout the day. These data are collected across multiple measurement scales, including continuous (physical activity), truncated (pain), ordinal (mood), and binary (daily life events). When indexed by time of day and stacked across assessment domains, these data structures can be
treated as multivariate functional data comprising continuous, truncated, ordinal, and binary variables. Motivated by
these applications, we propose a multivariate functional principal component analysis for mixed-type data
 (M$^2$FPCA). The approach is based on a semiparametric Gaussian copula model and assumes that the observed
data arise from an underlying multivariate generalized latent nonparanormal functional process. Latent temporal and
inter-variable dependence are estimated semiparametrically through Kendall’s tau bridging method. Two covariance
estimation procedures are developed: a fully multivariate block-wise estimator and a computationally efficient alternative
based on partial separability that assumes shared principal components across domains. The proposed method yields interpretable latent functional principal component scores that can serve as participant-specific digital biomarkers. Simulation studies demonstrate the method’s competitive performance under various complex dependence structures. The method is applied to mHealth data from 307 participants in the National Institute of Mental Health Family Study of Mood and Affective Spectrum Disorders.
Our approach identifies time-of-day patterns shared across mood, anxiety, energy, and physical activity that clinically meaningfully stratify mood disorder subtypes.
\end{abstract}

\section{Introduction}\label{sec:intro}
Advances in wearable and sensor technologies have enabled modern mobile health (mHealth) studies to collect repeated within-day measurements of participants’ mood, energy, stress, headache, and other health experiences using ecological momentary assessment (EMA) \citep{husky2025real}, alongside continuous behavioral measures such as sleep and physical activity from wearable devices \citep{dey2024functional, merikangas2019real}. These high-frequency observations provide rich opportunities to study temporal behavioral dynamics and their associations with health outcomes. However, the joint analysis of such data poses substantial statistical and computational challenges: observations may occur at irregular times, exhibit strong subject heterogeneity, and arise from mixed measurement scales including binary, ordinal, truncated, and continuous variables. Functional data analysis \citep{Ramsay05functionaldata,crainiceanu2024functional} provides a natural framework for studying processes observed over a continuum such as time. When indexed by time of day, multimodal EMA assessments together with wearable-derived activity measurements can be viewed as mixed-type multivariate functional data.
\subsection{Motivating Application}
\begin{figure}[htbp]
    \centering
    \includegraphics[width=\textwidth]{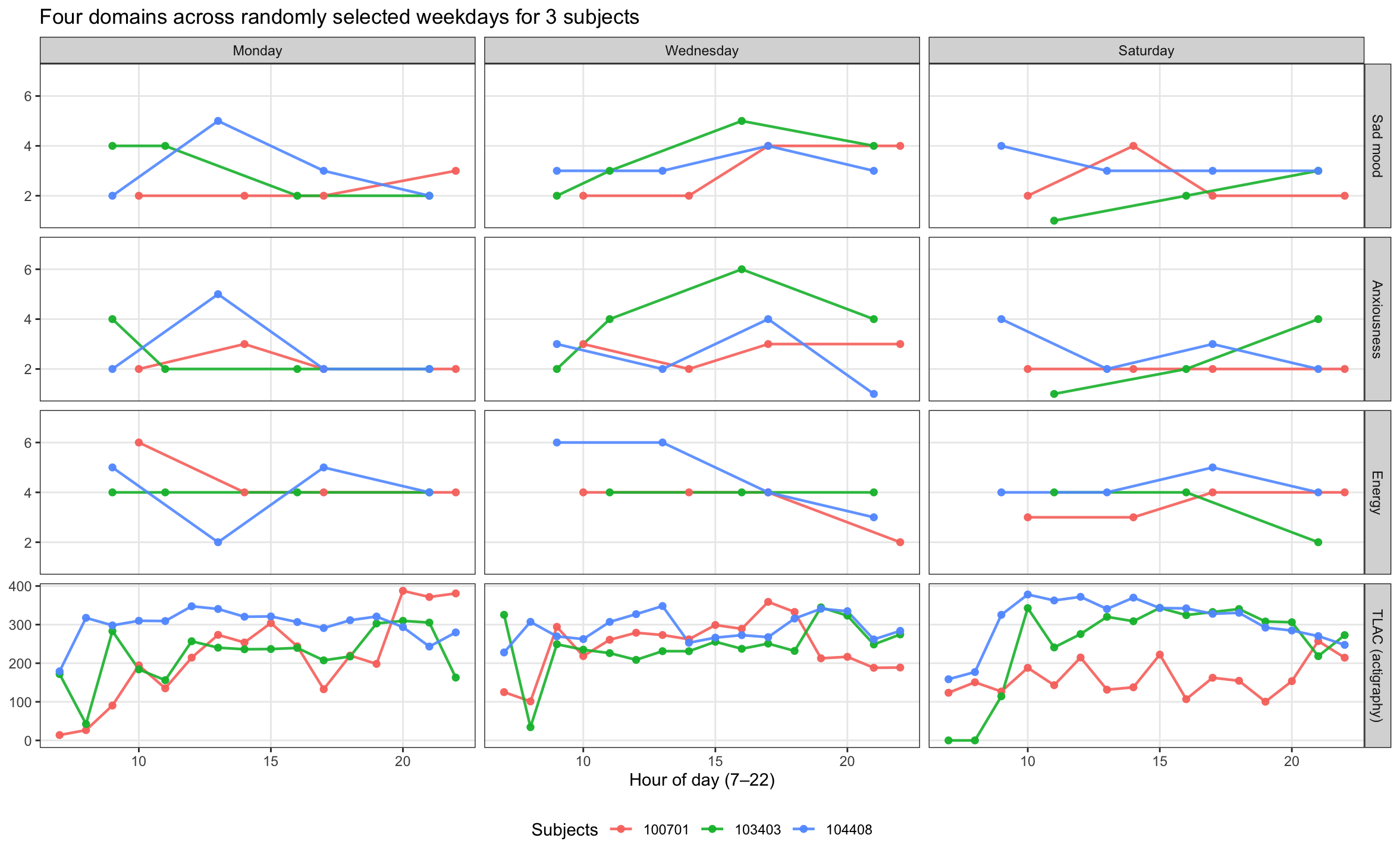}
    \caption{
    Diurnal trajectories across four behavioral domains for selected participants on three days of the week.
    Rows correspond to domains (sad mood, anxiousness, energy, and total locomotor activity), and columns correspond to days of the week.
    Within each panel, colored points represent observed subject-specific weekday-averaged measurements collected at irregular times between 7:00 and 22:00, while line segments connect consecutive observed time points for each subject.
    Ordinal EMA domains are displayed on a common scale from 1 to 7, whereas actigraphy retains its natural scale.
    The figure highlights substantial between-subject heterogeneity, irregular observation patterns, and cross-domain temporal structure that motivate a joint multivariate functional modeling approach.
    }
    \label{fig:motivating_weekday_domains}
\end{figure}

We consider multimodal behavioral data collected using EMA and wearable sensors from the National Institute of Mental Health Family Study of the Mood Disorder Spectrum. Figure~\ref{fig:motivating_weekday_domains} illustrates raw diurnal trajectories of mood, anxiousness, energy, and physical activity across selected days for several participants, highlighting irregular observation times, substantial between-subject heterogeneity, and complex cross-domain temporal dependence. Mood, anxiousness, and energy are self-reported on a $(1)$–$(7)$ Likert ordinal scale, while physical activity is continuously measured via actigraphy. Understanding the time-varying interdependence among such multimodal behavioral signals \citep{merikangas2019real, stapp2023specificity, lateef2024association, lamers2018mood} is central to many mHealth studies. In this application, our goal is to characterize the joint diurnal dependence between mood, anxiousness, energy, and physical activity \citep{kang2023integrative,glaus2023objectively}, and to investigate how differences in within-day dynamics across these domains relate to major mood disorder subtypes.

\subsection{Background and Related Works}

Several approaches in functional data analysis (FDA) address multivariate continuous functional data. Early multivariate functional principal component analysis (MFPCA) methods were developed for functions observed on a common one-dimensional domain \citep{berrendero2011principal,jacques2014functional,chiou2014multivariate}. More general formulations allowing functions defined on different domains were proposed by \cite{happ2018multivariate}, who established theoretical foundations using a multivariate Karhunen–Loève representation. Multivariate Gaussian process (GP) models \citep{banerjee2014hierarchical,qiao2019functional} have also been widely used for modeling multivariate functional data. Extensions include Gaussian copula-based models for skewed functional data \citep{staicu2012modeling,alam2024modeling}, functional Gaussian graphical models for conditional dependence \citep{zhu2016bayesian,zapata2022partial}, and graphical Gaussian process models for structured multivariate dependence \citep{dey2025graph}.  

While these approaches provide flexible tools for continuous functional data, modern mHealth studies often generate mixed-type functional observations, including binary, ordinal, truncated, and continuous measurements. Existing work on non-Gaussian functional data largely focuses on generalized functional regression models \citep{gertheiss2015marginal,goldsmith2015generalized,scheipl2016generalized,meyer2022ordinal} and computationally efficient generalized FPCA methods \citep{leroux2023fast}, as well as generalized GP regression for non-Gaussian functional responses \citep{wang2014generalized}. However, these approaches are primarily designed for regression settings and do not provide coherent joint models for multivariate mixed-type functional outcomes. 

Among joint modeling approaches, \cite{li2014hierarchical} proposed a hierarchical model for bivariate functional data consisting of one continuous and one binary variable, linking dependence through latent principal component scores. Copula-based approaches have also been developed for skewed functional data \citep{staicu2012modeling,alam2024modeling} and for multivariate functional graphical models \citep{solea2022copula}, although these methods are restricted to continuous data types.  

For mixed data in non-functional settings, latent semiparametric Gaussian copula (SGC) models \citep{liu2009nonparanormal,liu2012high,fan2017high,yoon2018sparse,dey2022semiparametric} provide flexible and scale-invariant frameworks that accommodate continuous, truncated, ordinal, and binary variables. In the functional setting, \citet{dey2024functional} proposed an FPCA method for univariate mixed-type functional data. That approach, however, is not designed for multivariate functional outcomes and therefore cannot characterize cross-covariation across functional components. Motivated by this limitation, we propose a unified multivariate FPCA framework for mixed-type multivariate functional data (hereafter, $M^2$-functional data). The resulting method, $M^2$FPCA, enables joint dimension reduction while accounting for both serial dependence within components and cross-dependence between components.

\subsection{Contribution of Proposed Work}
The key methodological contributions of this paper are as follows. \textbf{i)} We introduce a \emph{multivariate generalized latent nonparanormal process} that generates the observed multivariate mixed-type functional data. This formulation provides a unified probabilistic framework for jointly modeling continuous, truncated, ordinal, and binary functional measurements. \textbf{ii)} We estimate latent temporal and inter-variable dependence semiparametrically using Kendall’s $\tau$ bridging within a semiparametric Gaussian copula (SGC) framework. Two complementary covariance estimation strategies are developed. The first yields a fully multivariate estimator based on blockwise estimation of latent covariance and cross-covariance surfaces, which forms the basis of the proposed \emph{$M^2$FPCA} procedure. To improve computational scalability when the number of domains is moderate or large, we further introduce a computationally efficient alternative based on a \emph{partial separability} assumption \citep{zapata2022partial}, leading to the \emph{ps--$M^2$FPCA} method. \textbf{iii)} The proposed framework yields interpretable latent principal components and multivariate functional principal component scores, providing a parsimonious representation of multivariate mixed-type functional trajectories. These latent scores serve as candidate digital biomarkers summarizing coordinated dynamics across multiple behavioral domains.

The rest of this article is organized as follows. Section~2 presents the modeling framework and assumptions underlying the data-generating mechanism and introduces the proposed FPCA procedures ($M^2$FPCA and ps--$M^2$FPCA). Section~3 evaluates the finite-sample performance of the proposed methods through simulation studies. Section~4 applies the proposed framework to ecological momentary assessment data from 307 participants in the National Institute of Mental Health Family Study of Mood and Affective Spectrum Disorders, including mood, anxiousness, energy, and activity trajectories. The analysis identifies dominant latent time-varying components across domains and recovers clinically meaningful digital phenotypes associated with mood disorder subtypes. Section~5 concludes with discussion and potential extensions.

\section{Methods}\label{sec:methods}
\subsection{Data structure}\label{sec:data_structure}

In this paper, we consider multivariate functional data, each observed on a common compact domain 
$\calT\subset\mathbb{R}$, representing the different time-of-day in our real data application. We make a note that the methodology developed in this paper is general and can be applied  to multivariate functional data observed on different domains
. The observed object is a $J$-variate functional process
$X(\cdot)=\{X_{1}(\cdot),\ldots,X_{J}(\cdot)\}$ with $X_{j}\in L^{2}(\calT)$ for $j=1,\ldots,J$, and each subject $i=1,\ldots,n$ provides an i.i.d.\ replicate
$X_{i}(\cdot)=\{X_{i1}(\cdot),\ldots,X_{iJ}(\cdot)\}$. The multivariate trajectories may be of mixed-type, including binary, ordinal, truncated, or continuous outcomes. In finite samples, for each subject $i$ and component $j$, we observe the realizations of the processes on an irregular, subject and component-specific grid; $
\{(t_{ijr},X_{ij}(t_{ijr})):\ r=1,\ldots,m_{ij}\}, \qquad t_{ijr}\in\calT,$
so that different components and subjects need not share the same observation times. Let the pooled set of observed time points (grid points) be denoted by,
$\calT_{\mathrm{obs}}=\bigcup_{i=1}^{n}\bigcup_{j=1}^{J}\bigcup_{r=1}^{m_{ij}}\{t_{ijr}\}$.
\subsection{Multivariate generalized latent nonparanormal process}
\label{sec:mgl_npp}

We model mixed-type multivariate functional data through a latent copula-based hierarchy that extends the univariate generalized latent nonparanormal process (GLNPP) of \citet{dey2024functional}. A latent multivariate Gaussian process induces cross-component dependence, component-wise monotone transformations define a latent continuous process, and type-specific observation maps generate the observed mixed-scale data.
\begin{definition}[Latent multivariate Gaussian process]
\label{def:latent_gp}
Let $V(\cdot)=\{V_1(\cdot),\ldots,V_J(\cdot)\}$ be a mean-zero $J$-variate Gaussian process on a common compact domain $\calT$, with covariance and cross-covariance functions
$C_{jk}(s,t)=\mathrm{Cov}\{V_j(s),V_k(t)\}$ for $s,t\in\calT$ and $j,k\in\{1,\ldots,J\}$.
\label{eq:latent_cov_def}
\end{definition}

\begin{definition}[Multivariate latent nonparanormal functional process]
\label{def:latent_npn}
A $J$-variate latent continuous process $Z(\cdot)=\{Z_1(\cdot),\ldots,Z_J(\cdot)\}$ follows a multivariate latent nonparanormal functional model if there exist monotone functions $\{f_t:\mathbb{R}\to\mathbb{R}\}_{t\in\calT}$ such that, pointwise in $t$, $V_j(t)=f_t\{Z_j(t)\}$ for $j=1,\ldots,J$, where $V(\cdot)$ is the latent multivariate Gaussian process in Definition~\ref{def:latent_gp}.
\end{definition}

\begin{definition}[Multivariate generalized latent nonparanormal process (MGLNPP)]
\label{def:glnpp}
Let $Z(\cdot)=\{Z_1(\cdot),\ldots,Z_J(\cdot)\}$ be as in Definition~\ref{def:latent_npn}. The observed multivariate functional process $X(\cdot)=\{X_1(\cdot),\ldots,X_J(\cdot)\}$ is said to follow an MGLNPP if, for each component $j$ and time $t\in\calT$, $X_j(t)$ is generated from $Z_j(t)$ through a type-specific observation map:
\begin{equation}
\begin{split}
X_j(t) &= Z_j(t), \hspace{4.4cm} \text{(continuous)},\\
X_j(t) &= Z_j(t)\,I\{Z_j(t)>\Delta_j(t)\}, \hspace{2.15cm} \text{(truncated)},\\
X_j(t) &= \sum_{k=0}^{L_j-1} k\,I\{\Delta_{jk}(t)\le Z_j(t)<\Delta_{j,k+1}(t)\},\\
&\qquad -\infty=\Delta_{j0}(t)\le\cdots\le\Delta_{jL_j}(t)=\infty, \hspace{0.45cm} \text{(ordinal)},\\
X_j(t) &= I\{Z_j(t)>\Delta_j(t)\}, \hspace{3.55cm} \text{(binary)}.
\end{split}
\label{eq:obs_map}
\end{equation}
The cutoff functions $\bDelta(\cdot)$ determine the marginal distributions and observation types.
\end{definition}

\begin{remark}[Identifiability]
\label{rem:identifiability}
Under Definition~\ref{def:glnpp}, identifiability operates at three levels: the latent Gaussian process $V(\cdot)$, the latent nonparanormal process $Z(\cdot)$, and the observation maps in \eqref{eq:obs_map}. At the copula level, the dependence structure of $V(\cdot)$ is identifiable up to correlation scaling \citep{dey2024functional}; under the constraint $C_{jj}(t,t)=1$, the correlation kernel $C$ is identifiable, whereas marginal variances are not. At the marginal level, the transformations linking $Z(\cdot)$ and $V(\cdot)$ are identifiable only for continuous or truncated components. For binary and ordinal components, these effects are absorbed into the cutoffs $\bDelta(\cdot)$, so the cutoffs are identifiable only up to monotone transformation.
\end{remark}

\begin{remark}[Relation to latent Gaussian process models]
For purely discrete (binary/ordinal) components, the proposed MGLNPP reduces to a standard latent Gaussian process model with thresholding, since discrete observations cannot distinguish marginal transformations from cutoff shifts. When both continuous and discrete components are present, however, the MGLNPP class is strictly richer: allowing marginal transformations for continuous components yields dependence structures not representable by latent Gaussian process models alone. This is a key advantage for mixed-type multivariate functional data.
\end{remark}

\begin{remark}[Finite-dimensional implication under irregular sampling]
\label{rem:finite_npn}
Let $S_1,\ldots,S_J$ be arbitrary finite subsets of $\calT$, possibly differing across components, and write $R=\sum_{j=1}^J |S_j|$. Under the MGLNPP model, the stacked latent vector $\big(Z_1(S_1)^\top,\ldots,Z_J(S_J)^\top\big)^\top$ follows a nonparanormal distribution with Gaussian copula determined by the restriction of the correlation kernel $C(\cdot,\cdot)$ to $(S_1,\ldots,S_J)$. This links the process-level formulation to irregular, subject-specific observation grids.
\end{remark}

\begin{figure}[htbp]
\centering
\begin{tikzpicture}[
  node distance=2.8cm,
  every node/.style={
    draw,
    rectangle,
    rounded corners,
    align=center,
    minimum width=6.5cm,
    minimum height=1.4cm,
    font=\small
  },
  arrow/.style={->, thick}
]

\node (V) {
$\boldsymbol{V}(\cdot) = (V_1,\ldots,V_J)$\\[1mm]
\textbf{Latent multivariate Gaussian process}\\
$\mathrm{Cov}\{V_j(s),V_k(t)\} = C_{jk}(s,t)$
};

\node (Z) [below of=V] {
$\boldsymbol{Z}(\cdot) = (Z_1,\ldots,Z_J)$\\[1mm]
\textbf{Latent continuous process}\\
$Z_j(t) = f_t^{-1}\{V_j(t)\}$
};

\node (X) [below of=Z] {
$\boldsymbol{X}(\cdot) = (X_1,\ldots,X_J)$\\[1mm]
\textbf{Observed mixed-type functional process}\\
Continuous \;|\; Truncated \;|\; Ordinal \;|\; Binary
};

\node (Obs) [below of=X] {
\textbf{Observed data}\\
$\{(t_{ijr}, X_{ij}(t_{ijr}))\}$\\
Irregular, subject- and component-specific grids
};

\draw[arrow] (V) -- node[right=6mm, draw=none, font=\small]
{Monotone marginal\\ transformations $f_t^{-1}$} (Z);

\draw[arrow] (Z) -- node[right=6mm, draw=none, font=\small]
{Type-specific thresholding\\ or truncation} (X);

\draw[arrow] (X) -- node[right=6mm, draw=none, font=\small]
{Irregular sampling} (Obs);

\end{tikzpicture}

\caption{Data-generating mechanism under the Multivariate Generalized Latent Nonparanormal Process (MGLNPP).
Cross-component dependence is induced through a latent multivariate Gaussian process.
Component-wise monotone transformations produce a latent continuous process, which is mapped to observed
mixed-type functional data via type-specific thresholding or truncation and observed on irregular grids.}
\label{fig:mgl_npp_flowchart}
\end{figure}
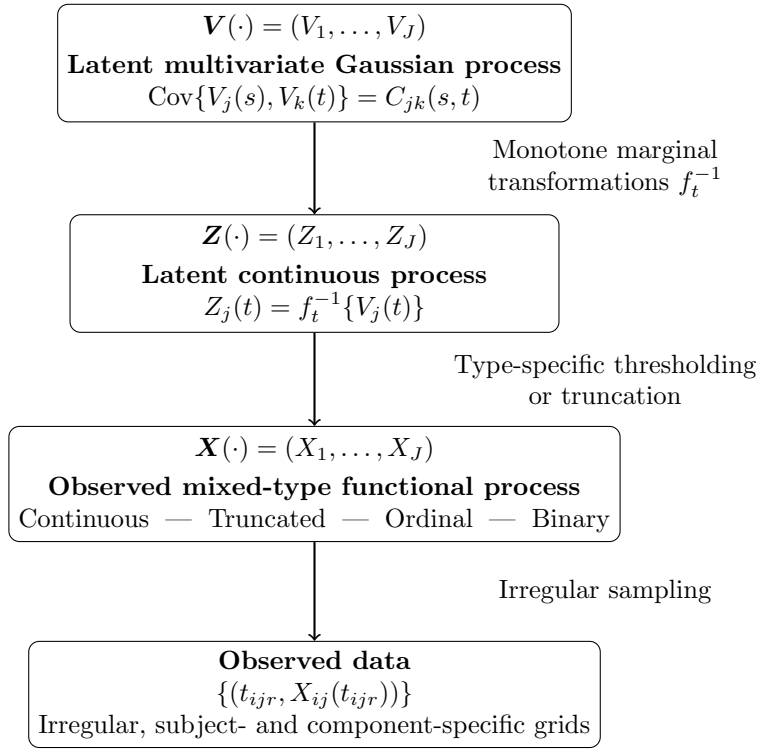

\paragraph{Assumption (Data-generating mechanism)}
\label{assump:mgl_npp}
We assume the observed multivariate functional data arise from an MGLNPP model consisting of (i) a latent $J$-variate Gaussian process with correlation kernel $C$, (ii) component-wise monotone marginal transformations, and (iii) type-specific thresholding or truncation maps. Subjects provide independent replicates observed on irregular, component-specific grids; Figure~\ref{fig:mgl_npp_flowchart} summarizes this construction.
\subsection{Estimation overview}\label{sec:est_overview}

Our primary target is the latent correlation kernel
$C=(C_{jk})$, which characterizes second-order dependence in the latent multivariate Gaussian process underlying the MGLNPP model. Because the observed processes $X_j(\cdot)$ may be discrete, truncated, or non-Gaussian, direct moment-based covariance estimation is not appropriate. Instead, we exploit the invariance of Kendall’s $\tau$ to monotone transformations and use the type-specific bridge, $
\tau_{jk}(s,t)=F_{jk}\{C_{jk}(s,t)\},$
where $F_{jk}$ depends on the marginal data types and cutoff structures; see \citet{dey2022semiparametric}, \citet{yoon2018sparse}, and \citet{dey2024functional}.

We denote the univariate GLNPP-based procedure of \citet{dey2024functional} by \emph{$M^1$FPCA}. Building on this, we propose two multivariate estimators. The first, \emph{$M^2$FPCA}, jointly estimates all marginal and cross-correlation surfaces $\{C_{jk}\}_{j,k=1}^J$ and then applies multivariate FPCA to the predicted latent Gaussian trajectories. The second, \emph{ps--$M^2$FPCA}, is a computationally scalable alternative that combines componentwise $M^1$FPCA latent predictions with a partially separable multivariate FPCA decomposition \citep{zapata2022partial}. We describe both procedures below. Additional nuanced implementation details are deferred to Section~\ref{suppl:estimation_tools} of the Appendix, including the bridging functions, estimation of cutoff parameters and monotone transformations, and selection of tuning parameters. 

\subsubsection{$M^2$FPCA: blockwise estimation and multivariate FPCA}
\label{sec:m2fpca}

The $M^2$FPCA procedure estimates the full latent correlation structure $\{C_{jk}\}_{j,k=1}^J$ blockwise and then performs multivariate FPCA on the predicted latent Gaussian trajectories.

\paragraph{Blockwise estimation}
For $(j,k)$ and $(s,t)\in\calT_{\mathrm{obs}}^2$, let $n_{jk}^{ii'}(s,t)=1$ if subjects $i$ and $i'$ are both observed for component $j$ at $s$ and component $k$ at $t$, and $0$ otherwise, and define $N_{jk}(s,t)=\sum_{1\le i<i'\le n} n_{jk}^{ii'}(s,t)$. The weighted sample Kendall statistic is
\begin{equation}\label{eq:kendall_hat_generic}
\hat{\tau}_{jk}(s,t)=\frac{2}{N_{jk}(s,t)}
\sum_{1\le i<i'\le n:\, n_{jk}^{ii'}(s,t)>0}
\mathrm{sgn}\!\big\{(X_{ij}(s)-X_{i'j}(s))(X_{ik}(t)-X_{i'k}(t))\big\},
\end{equation}
whenever $N_{jk}(s,t)>0$. We treat $N_{jk}(s,t)$ as a reliability weight and exclude pairs with $N_{jk}(s,t)\le c_0$.

Each latent correlation surface is modeled by the tensor-product spline representation
\begin{equation}
C_{jk}(s,t)=g\!\left(\sum_{a=1}^{K}\sum_{b=1}^{K} u_{ab}^{(jk)} B_a(s)B_b(t)\right),
\qquad
g(x)=\frac{e^x-1}{e^x+1},
\label{eq:spline_corr_model}
\end{equation}
with symmetry constraints for marginal surfaces ($j=k$). The spline coefficients are estimated by weighted nonlinear least squares:
\begin{equation}
\widehat{\mathbf U}^{(jk)}
=
\arg\min_{\mathbf U^{(jk)}}
\sum_{(s,t)\in\calT_{\mathrm{obs}}^2:\,N_{jk}(s,t)>c_0}
N_{jk}(s,t)
\left[
\hat\tau_{jk}(s,t)-F_{jk}\!\left(
g\!\left(\sum_{a=1}^{K}\sum_{b=1}^{K} u_{ab}^{(jk)}B_a(s)B_b(t)\right)
\right)
\right]^2 .
\label{eq:wls_generic}
\end{equation}

\paragraph{Latent prediction and multivariate FPCA}
Evaluating $\widehat C_{jk}$ on a grid $S=\{t_1,\ldots,t_m\}$ yields blocks $\widehat\Sigma_{jk}\in\mathbb{R}^{m\times m}$, which are assembled into the covariance operator $\widehat{\mathcal C}$. We enforce positive definiteness by eigenvalue thresholding: if $\widehat{\mathcal C}=P\Lambda P^\top$, define $\widehat{\mathcal C}_{\mathrm{PD}}=P\Lambda'P^\top$ with $\Lambda'_{pp}=\max(\Lambda_{pp},\epsilon)$. Let $\bV_i(t)=\sum_{\ell\ge1}\xi_{i\ell}\bphi_\ell(t)$ denote the multivariate Karhunen--Lo\`eve expansion of the latent Gaussian process. We obtain subject-specific latent predictions by the BLUP -- $
\widehat{\bV}_i(t)=\mathbb{E}\!\left\{\bV_i(t)\mid \bX_i,\widehat{\mathcal C}_{\mathrm{PD}}\right\},$
and then apply multivariate FPCA to $\{\widehat{\bV}_i\}_{i=1}^n$ to obtain estimated eigenfunctions and scores \citep{happ2018multivariate, MFPCApackage}. Algorithm~\ref{alg:sg_mfpca_condensed} in the  Appendix summarizes the procedure.

\paragraph{Computational complexity}
The dominant cost is $O(J^2|S|^2 n\log n)$ for blockwise Kendall estimation and $O(J^3|S|^3)$ for latent prediction and multivariate FPCA.

\subsubsection{ps--$M^2$FPCA via $M^1$FPCA latent predictions}
\label{sec:ps_sg_mfpca}

While $M^2$FPCA flexibly estimates the full latent dependence structure, its cubic scaling in $J|S|$ can be burdensome when $J$ or $|S|$ is moderate to large. We therefore consider a scalable alternative, \emph{ps--$M^2$FPCA}, based on a partially separable representation \citep{zapata2022partial}:
\begin{equation}
\bV_i(t)=\sum_{\ell=1}^{\infty}\boldsymbol{\bxi}_{i\ell}\phi_\ell(t),
\qquad
\boldsymbol{\bxi}_{i\ell}=(\bxi_{i\ell1},\ldots,\bxi_{i\ell J})^\top,
\label{eq:ps_kl}
\end{equation}
where $\{\phi_\ell\}$ are temporal eigenfunctions shared across components.

The procedure has two stages. In Stage I, we apply $M^1$FPCA separately to each component to estimate $C_{jj}$ and obtain univariate latent predictions $\widehat V_{ij}(t)=\mathbb{E}\{V_{ij}(t)\mid \bX_{ij},\widehat C_{jj}\}$ on a common grid $S$. In Stage II, we apply the partially separable MFPCA construction of \citet{zapata2022partial} to the multivariate latent predictions $\widehat{\bV}_i(t)=(\widehat V_{i1}(t),\ldots,\widehat V_{iJ}(t))^\top$, estimating shared temporal eigenfunctions from the pooled marginal covariance $H(t,t')=J^{-1}\sum_{j=1}^J \widehat C_{jj}(t,t')$ and recovering component-specific score covariance matrices by projection. A detailed implementation is provided in Algorithm~\ref{alg:ps_sg_fpca} in the Appendix. The resulting computational cost is $O(J|S|^2 n\log n)+O(J|S|^3)+O(|S|^3)$, compared with $O(J^2|S|^2 n\log n)+O(J^3|S|^3)$ for $M^2$FPCA. Thus ps--$M^2$FPCA substantially reduces the dominant cubic cost when $J$ is moderate or large.

\subsection{Multivariate curve prediction via latent BLUP}\label{sec:curve_prediction}

Let $X_i(\cdot)=\{X_{i1}(\cdot),\ldots,X_{iJ}(\cdot)\}$ be observed at irregular, component-specific times and suppose prediction is desired at new time points. Let $\bV_i^O$ and $\bV_i^N$ denote the corresponding observed and new latent Gaussian vectors. Under the fitted multivariate latent Gaussian model, $\big((\bV_i^O)^T,(\bV_i^N)^T\big)^T$ is jointly Gaussian with block covariance matrices $\bC^{O,O}$, $\bC^{O,N}$, $\bC^{N,O}$, and $\bC^{N,N}$ assembled from the estimated latent cross-covariance surfaces $\{\widehat C_{jk}\}_{j,k=1}^J$ evaluated at the relevant time points. Prediction is therefore based on the full multivariate latent vector rather than on separate components, with conditional mean $E(\bV_i^N\mid\bV_i^O)=\bC^{N,O}(\bC^{O,O})^{-1}\bV_i^O$ and conditional covariance $\bC^{N,N}-\bC^{N,O}(\bC^{O,O})^{-1}\bC^{O,N}$. The resulting multivariate BLUP is $\widehat{\bV}_i^N=\widehat{\bC}^{N,O}(\widehat{\bC}^{O,O})^{-1}\widehat{\bV}_i^O$, where $\widehat{\bV}_i^O=E(\bV_i^O\mid X_i^O)$ is computed using the multivariate conditional expectation approach of \citet{dey2022semiparametric}. Unlike the univariate BLUP in \citet{dey2024functional}, this prediction borrows strength across domains through the full latent covariance structure. Predicted trajectories on the observed scale are then obtained using the estimated cutoffs and, when needed, transformation functions; for a binary component, $\widehat X_{ij}^N(t)=I\{\widehat V_{ij}^N(t)>\widehat\Delta_j(t)\}$.

\section{Simulation Study}

\subsection{Data Generating Scenarios}
We investigate the performance of the proposed methods, $M^2$FPCA and ps-$M^2$FPCA, via numerical simulations. We consider a four-dimensional ($p=4$) observed multivariate functional process $X(\cdot) = (X^{(b)}(\cdot), X^{(o)}(\cdot), X^{(tr)}(\cdot), X^{(c)}(\cdot))^\top$, defined on a domain $\mathcal{T}=[0,1]$. The functional processes $X^{(type)}(\cdot), type\in{b,o,tr,c}$ (binary, ordinal, truncated, continuous) are observed on a regular grid of $m=16$ equidistant points. We vary the sample size $n \in \{100, 500, 1000\}$ to assess the method's performance across different sample sizes. The observed multivariate mixed-type data follow a multivariate generalized latent nonparanormal process (MGLNPP) and are generated based on a latent multivariate Gaussian process $V(\cdot)$ as described below.

\noindent \textbf{Stationary Covariance:}
In the first scenario, the latent process $V(\cdot)=\{V_1(\cdot),\,V_2(\cdot),V_3(\cdot),V_4(\cdot)\}$ is generated from a stationary zero-mean multivariate Gaussian process (GP) with a multivariate Matérn cross-covariance function. The covariance between the $i$-th variable at time $s$ and the $j$-th variable at time $t$ is defined as, $
C_{ij}(s,t) = \sigma_{ij} \exp\left(-\phi_{ij} |s-t|\right)$
This corresponds to a multivariate Matérn covariance structure with smoothness parameters fixed at $\nu_{ij} = 0.5$ for all $i,j$ (the exponential kernel). The marginal length-scale parameters $\{\phi_{ii}\}_{i=1}^p$ are generated randomly permuting a sequence of $p$ equidistant values spanning the interval $[1, 5]$. To ensure the validity of the multivariate covariance function, the cross-length-scale parameters $\phi_{ij}$ (for $i \neq j$) are defined as the square root of the average of the squared marginal length-scales:
$
\phi_{ij} = \sqrt{\frac{\phi_{ii}^2 + \phi_{jj}^2}{2}}
$. The base correlation structure is governed by an exchangeable correlation matrix $\mathbf{R}$ with $R_{ij} = 1$ for $i=j$ and $R_{ij} = 0.5$ for $i \ne j$.
Finally, the cross-covariance parameters $\sigma_{ij}$ are derived to ensure the multivariate Matérn covariance matrix is positive definite. With marginal variances set to $\sigma_{ii} = 1$, the cross-covariance parameters $\sigma_{ij}$ are calculated as:
$
\sigma_{ij} = R_{ij} \frac{\sqrt{\phi_{ii} \phi_{jj}}}{\phi_{ij}}$
This specification ensures that the resulting cross-covariance functions yield a valid, positive-definite multivariate covariance operator.


\noindent \textbf{Non-stationary Covariance:}
In the second scenario, we generate the latent process $V(\cdot)$ using a basis expansion approach to induce non-stationarity.The process is defined as $V_{i}(t) = \sum_{l=1}^{M} \bm\theta_{il} \phi_l(t)$, where $\{\phi_l(t)\}$ are Fourier basis functions on $[0,1]$ and $M=101$ is the number of basis functions. The basis coefficients $\bm\theta_{il} = (\theta_{i1l}, \dots, \theta_{i4l})^\top$ are generated independently from  $N_4(0, \*\Sigma_l)$. The covariance matrices were set to $\*\Sigma_l=a_l\Omega_l^{-1}$ \citep{dey2025graph}, with $ a_l=3l^{-1.8}$. The precision matrices $\Omega_l$ are generated using the algorithm in \cite{peng2009partial} as dense matrices, assuming a complete graph structure among the four variables, thereby inducing dependence between all variable pairs. This scenario introduces a partially separable covariance structure between the functional variables $\{V_1(\cdot),V_2(\cdot),V_3(\cdot),V_4(\cdot)\}$.

\noindent \textbf{Generation of Mixed-Type Data:}
The observed mixed-type functional data $X(\cdot) = (X^{(b)}(\cdot), X^{(o)}(\cdot), X^{(t)}(\cdot), X^{(c)}(\cdot))^\top$ are obtained by transforming the standardized latent process $V(\cdot)$ (scaled to have unit variance) through element-wise thresholding and truncation operations. In the simulations, we consider the special case where the monotone marginal transformation is the identity, so the observed mixed-type data are generated directly from the latent Gaussian process. The four variables are generated as follows:

\begin{itemize}
    \item \textbf{Binary ($X^{(b)}$):} The first component is generated by thresholding the latent process at $\delta_b = 0.5$, such that $X^{(b)}_{i}(t) = I(V_{i1}(t) > 0.5)$.
    \item \textbf{Ordinal ($X^{(o)}$):} The second component is discretized into ordinal categories using cutoffs $\delta = (-\infty, -0.6, 0.1, 0.6, \infty)$. The observed value is determined by the interval in which the latent value $V_{i2}(t)$ falls, resulting in four ordinal levels $0,1,2,3$.
    \item \textbf{Truncated ($X^{(tr)}$):} The third component represents a zero-truncated continuous variable. It retains the value of the latent process only if it exceeds a threshold $\delta_b = 0.5$, otherwise it is set to zero: $X^{(tr)}_{i}(t) = V_{i3}(t) \cdot I(V_{i3}(t) > 0.5)$.
    \item \textbf{Continuous ($X^{(c)}$):} The fourth component is considered to be the continuous latent process: $X^{(c)}_{i}(t) = V_{i4}(t)$.
\end{itemize}

\subsection{Simulation Results}
We used 100 Monte-Carlo (M.C) replications across the scenarios presented above (two correlation kernel and three sample sizes ) to assess the performance of the proposed $M^2$FPCA and ps--$M^2$FPCA methods. As an competing approach, we used the MFPCA \citep{happ2018multivariate} method developed for continuous multivariate functional data on the observed data $X(\cdot) = (X^{(b)}(\cdot), X^{(o)}(\cdot), X^{(tr)}(\cdot), X^{(c)}(\cdot))^\top$. The performance of the competing methods are assessed in terms of the integrated square errors for estimation of the covariance functions, where ISE is defined as $ISE=\int_0^1\int_0^1\{C_T(s,t)-\hat{C}(s,t)\}^2dsdt$, where $C_T(s,t)$ is the true covariance function and $\hat{C}(s,t)$ is the estimated covariance. The ISE  values of the estimators across the different correlation kernel types and sample-sizes are reported in Table \ref{tab:sim-table-new}. 
\begin{table}[H]
\caption{Average ISE (and corresponding sd) between true and estimated covariance matrices for different methods and sample sizes. The minimum average ISE is presented in bold-face.}
\label{tab:sim-table-new}
\resizebox{\textwidth}{!}{\color{black}{\begin{tabular}{@{}l|lll|lll@{}}
\toprule
Sample Size & \multicolumn{3}{c|}{Stationary} & \multicolumn{3}{c}{Non-stationary} \\ \cmidrule(l){2-7} 
 & $M^2$FPCA & ps-$M^2$FPCA & MFPCA\_N & $M^2$FPCA & ps-$M^2$FPCA & MFPCA\_N \\ \hline
$n=100$ &0.011 (0.002)  &0.031 (0.05)  &0.042 (0.003)  &0.00094 (0.002)  &0.00092 (0.023)  &0.030 (0.002)  \\ \hline
$n=500$ &0.003 (0.0004)  &0.024 (0.047)  &0.038 (0.001)  &0.0024 (0.0004)  &0.0025 (0.0005)  &0.028 (0.0006)  \\ \hline
$n=1000$ &0.002 (0.0002)  &0.017 (0.033)  &0.038 (0.0008)  &0.0015 (0.0002)  &0.0021 (0.0003)  &0.028 (0.0005)  \\ \bottomrule
\end{tabular}}}
\end{table}

\begin{figure}[H]
\centering
\includegraphics[width=0.65\linewidth , height=0.8\linewidth]{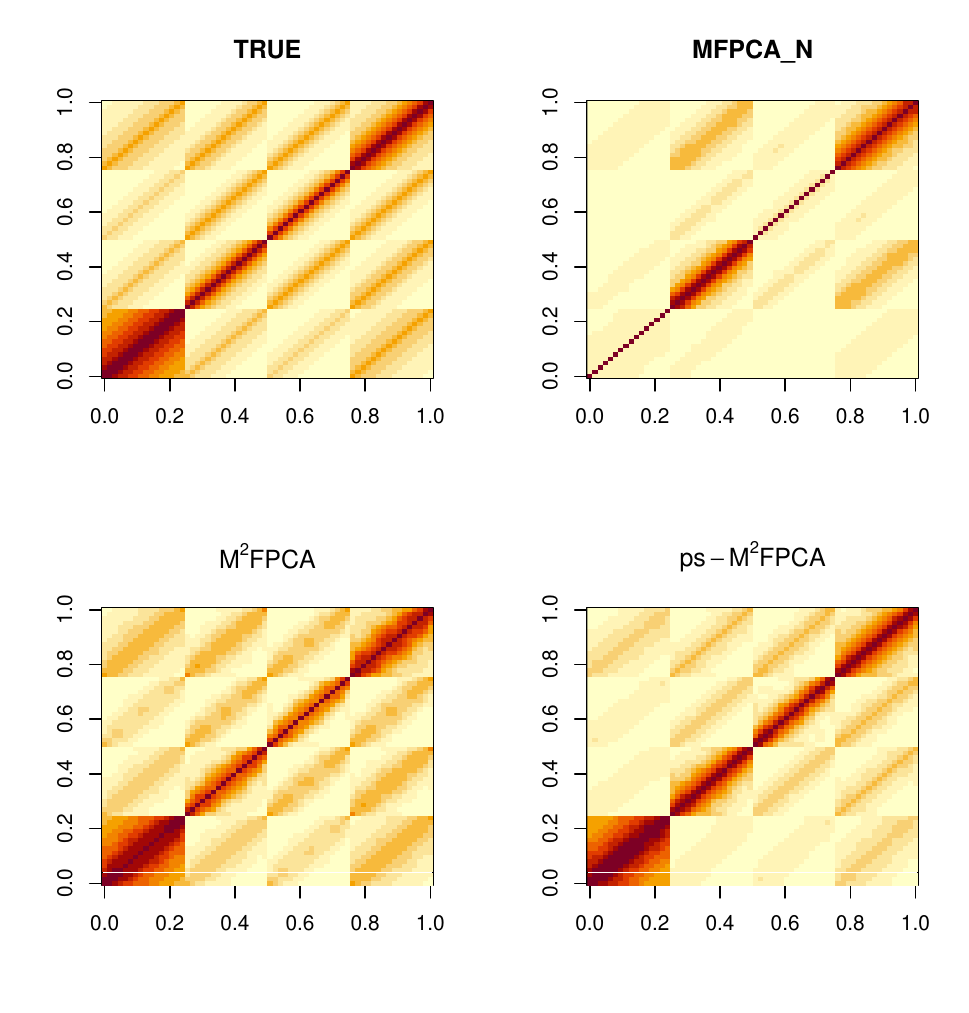}
\caption{True and the Monte-Carlo mean of the estimated covariance surface for the stationary correlation kernel and $n=500$, from the naive MFPCA (MFPCA$_N$) and the proposed $M^2$FPCA, ps-$M^2$FPCA.}
\label{fig:fig3new2}
\end{figure}

It can be observed for the stationary covariance case (which is non-separable), the proposed $M^2$FPCA approach provides approximately 6 times smaller ISE (on average) compared to ps--$M^2$FPCA and 12 times smaller ISE (on average) compared to MFPCA.
As sample sizes increase, the ISEs from both $M^2$FPCA and ps--$M^2$FPCA decreases, indicating a satisfactory performance of these methods. We display in Figure \ref{fig:fig3new2} the Monte-Carlo mean of the estimated covariance surfaces for sample size $n=500$. Both $M^2$FPCA and ps--$M^2$FPCA can be seen closely capturing the true correlation kernel. The estimates for sample size $n=100,1000$ are reported in Figures \ref{fig:fig5new1}, \ref{fig:fig5new2} of the Appendix.


For the non-stationary covariance case, which results into a partially separable covariance structure, the performance of the proposed $M^2$FPCA and ps--$M^2$FPCA (which assumes partial separability) are very similar. On an average both the methods provide approximately 20 times smaller ISE compared to MFPCA, with perforamnces getting better as sample size increase. We display in Figure \ref{fig:fig4new1} the Monte-Carlo mean of the estimated covariance surfaces for sample size $n=500$. Both the $M^2$FPCA and ps--$M^2$FPCA can be seen closely capturing the true non-stationary correlation kernel. The estimates for sample size $n=100,1000$ are reported in Figures \ref{fig:fig6new1}, \ref{fig:fig6new2}  of the Appendix.




\section{Application to NIMH Family Study}
\noindent
We analyzed EMA data from the National Institute of Mental Health (NIMH) Family Study of Mood and Affective Spectrum Disorders \citep{merikangas2014independence}, a community-based cohort recruited from the greater Washington, DC, metropolitan area. Participants ($N=307$; ages 11--85 years) represented a full diagnostic spectrum: controls with no lifetime mood disorder ($130$), major depressive disorder (MDD; $106$), bipolar I disorder ($41$), and bipolar II disorder ($30$). All procedures were approved by the NIH Combined Neuroscience Institutional Review Board, and informed consent was obtained from all participants. Supplementary Table \ref{tab:table1_baseline} reports the descriptive summary of the sample.

\smallskip

Each participant wore a wrist-worn actigraphy monitor (Philips Respironics Actiwatch) continuously for 14 consecutive days, yielding minute-level activity counts. Actigraphy data were processed using R package \texttt{GGIR} \citep{migueles2019ggir, guo2022processing, GGIR2025}. EMA prompts were administered four times daily through hand-held digital devices over 14~days, scheduled relative to each participant’s individual wake--sleep cycle to maximize ecological validity \citep{Hall2021, deVries2021}. Participants rated their current sad mood, anxiousness, and energy on ordinal Likert scales ($1$ (very low) – $7$ (very high)); due to sparsity in the upper range, categories $6,$ and $7$ were collapsed into a single category ($6$) to ensure stable estimation. Although each day contributes only four EMA prompts, the effective temporal information is richer than four fixed design points because prompts were scheduled relative to each participant’s wake--sleep cycle and therefore occurred at different clock times across participants and days. After alignment to a common 7{:}00~a.m.--10{:}00~p.m.\ waking window, these irregularly located observations jointly sample the underlying diurnal trajectories over the full domain. Our framework pools information across all subject-days and borrows strength from the densely observed actigraphy process. This estimation permits recovery of more than four interpretable modes of variation beyond what would be identifiable from four fixed common assessment times alone. 

To obtain harmonized functional trajectories, all observations were aggregated into 16 equal one-hour bins. Actigraphy signals were log-transformed and summed within each bin to compute bin-specific Total Log Activity Count (TLAC). EMA responses within a bin were averaged and then rounded to the nearest ordinal level to preserve categorical interpretation. This produced a four-dimensional daily function for each participant-day,
\[
\big(\mathrm{Sad\ Mood}(t),\ \mathrm{Anxiousness}(t),\ \mathrm{Energy}(t),\ \mathrm{TLAC}(t)\big),\quad t=1,\dots,16.
\]
The Family Study achieved a high average EMA completion rate of $93\%$ (range: $83-98\%$) with minimal fatigue effects and an average response duration of $3.3$~minutes. Missing EMA or actigraphy observations were balanced across groups, and not associated with time of day or participant attributes. Missing entries were thus treated as missing at random. Across all participants, the analysis included $3474$ subject-days meeting a minimum wear and response-time criterion.

\smallskip

\begin{figure}[htbp]
    \centering
    \includegraphics[width=0.7\textwidth]{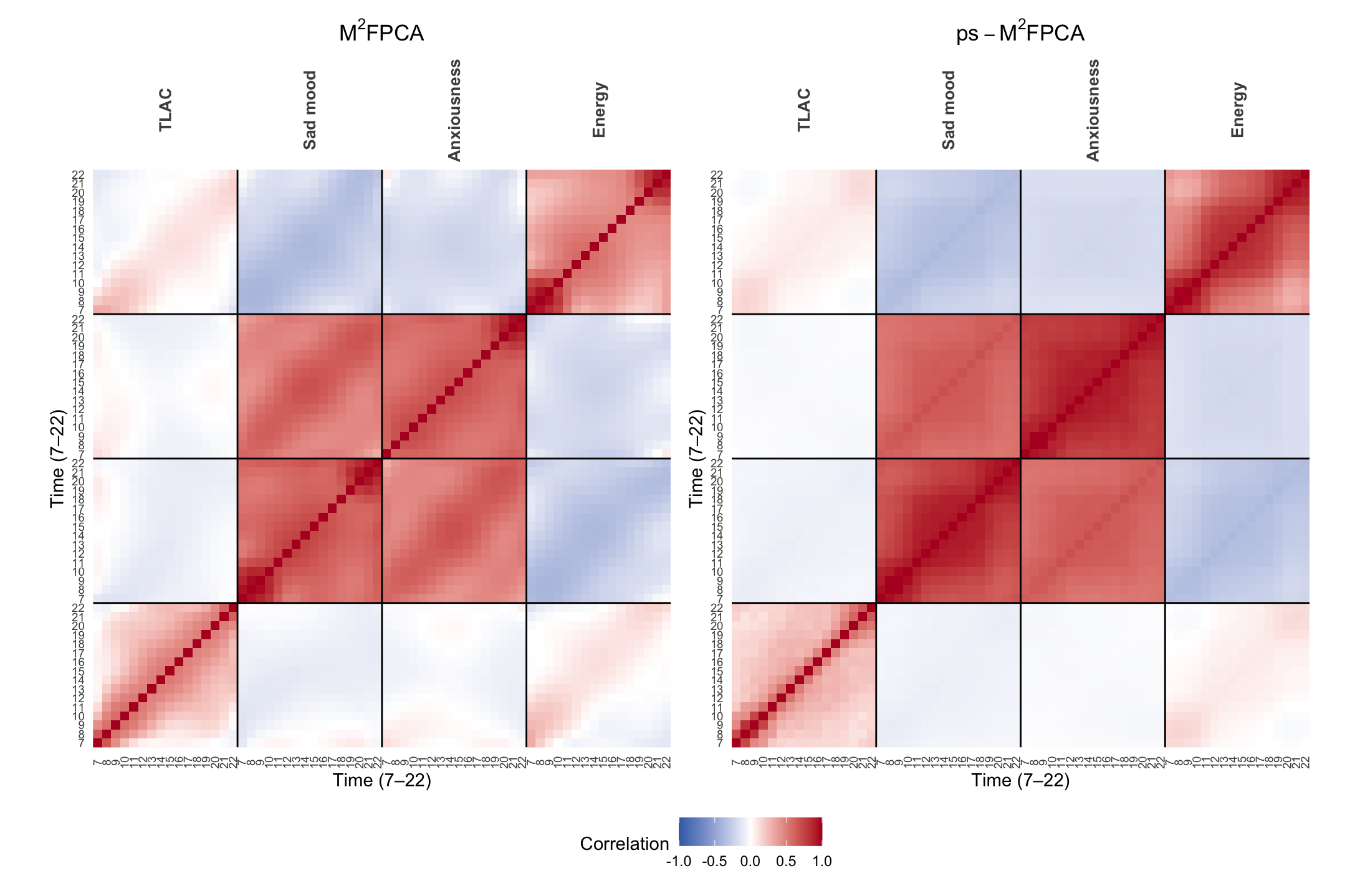}
    \caption{Comparison of estimated correlation surfaces under the fully multivariate $M^2$FPCA and partially separable (PS) models.}
    \label{fig:corr_comp}
\end{figure}

The primary aim of our data-analysis is to characterize the joint time-of-day (diurnal) dependence between mood,
anxiousness, energy, and TLAC, and to quantify how differences in within-day dynamics across these domains are associated
with major mood disorder subtypes. First, we applied $M^2$FPCA and ps-$M^2$FPCA to estimate corresponding latent multivariate correlation surfaces. The estimated surfaces from both  approaches are shown in Figure \ref{fig:corr_comp} and can be seen to capture similar time-of-day dependence within and between the domains.

In particular, we observe that higher TLAC is inversely associated with sad mood and anxiousness during most of the day, while TLAC is positively associated with energy. Interestingly, there is a negative association between lagged TLAC and energy, particularly at higher lags between the observations. Sad mood and anxiousness are found to be highly positively correlated (Fig. \ref{fig:corr_comp}). A strong negative association is observed between the sad mood and energy and anxiousness and energy. 
To assess the plausibility of the partial separability assumption, following the strategy of \citep{zapata2022partial}, we conducted a series of diagnostic comparisons. Specifically, we compared FPCs from ps-$M^2$FPCA and $M^1$FPCA, and examined the covariance structure of the univariate scores against the covariance implied by the ps-$M^2$FPCA scores. Visual inspection revealed no meaningful deviations between these quantities (Figures~\ref{fig:eigen_comp} and~\ref{fig:cov_comp}), suggesting that the partial separability assumption does not materially distort the inter-domain dependence in our data. Since ps-$M^2$FPCA is much more computationally tractable and provides more straightforward clinical interpretability of domain-specific latent PC scores, we focus on ps-$M^2$FPCA in the remainder of this section.



The eigenfunctions from ps-$M^2$FPCA are shown in Figure \ref{fig:eigen_comp} and reveal shared latent diurnal processes across the four domains. Under ps-$M^2$FPCA, the top three latent components explained $82\%$ of the total variability. Specifically, FPC1 explained $64\%$ of the variability and primarily reflected the overall daily burden of symptoms and activity. FPC2 explained $12\%$ of the variability and captured a monotone morning-to-evening contrast across all four domains, consistent with a homeostatic process in which sleep pressure accumulates with time since waking. FPC3 explained $6\%$ of the variability and captured an inverted-U pattern peaking around midday, suggestive of a circadian influence on the timing of affective and activation states. 

Since the top two components together explained $76\%$ of the total variability, we focused on FPC1 and FPC2 in the subsequent analysis. For each subject-day and domain, we estimated latent functional principal component scores (FPC scores), and then computed subject-level means and standard deviations over the study period. This yielded $4\times 2\times 2=16$ clinically interpretable digital biomarkers per participant (four domains $\times$ two FPC scores  $\times$ two summaries: mean, SD). To evaluate discriminative power, we used these subject-level digital biomarkers in a multinomial logistic regression model to distinguish Bipolar I ($41$), Bipolar II ($30$), and MDD ($106$) from healthy controls ($130$). Because the feature space was moderately high-dimensional with $16$ candidate predictors relative to the available sample size of 307 subjects, LASSO regularization was employed to perform variable selection to identify important biomarkers while reducing overfitting. 

\begin{figure}[!t]
\centering
\includegraphics[width=0.9\textwidth]{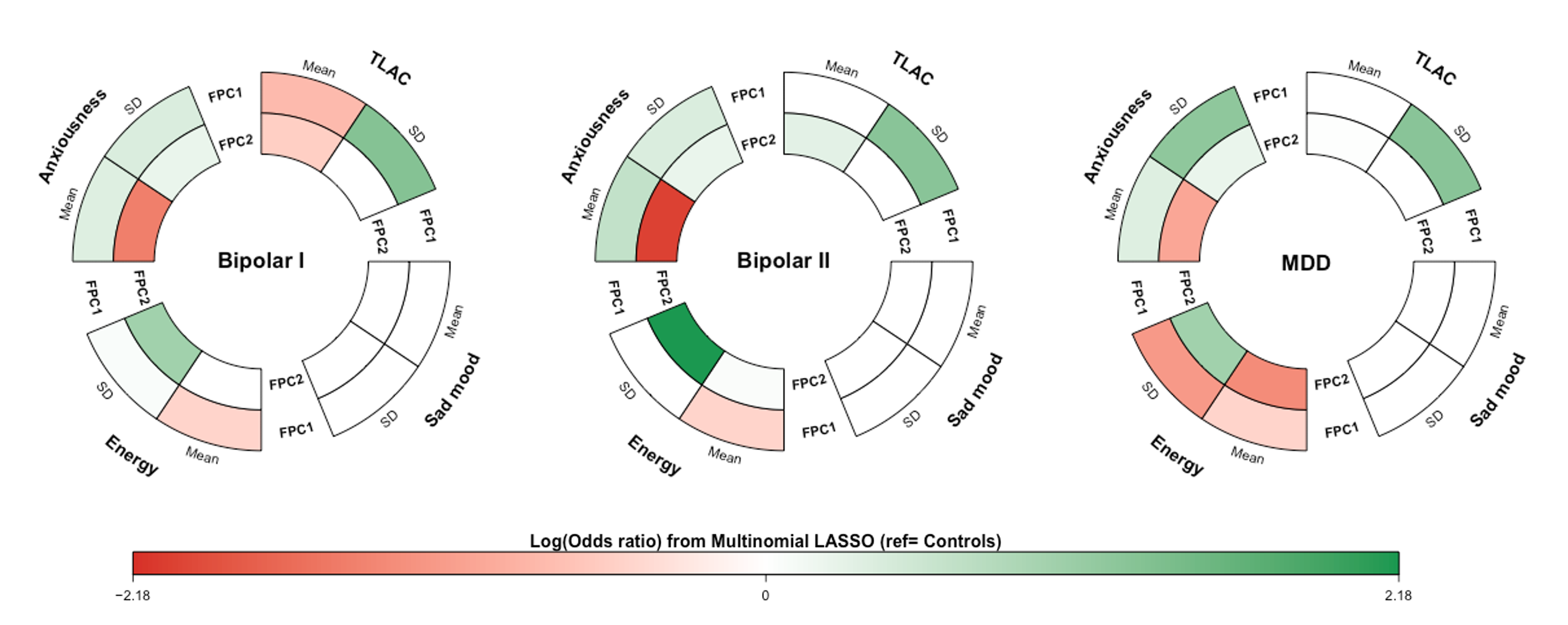}
\caption{Effects of latent daily features in a multinomial logistic LASSO model classifying diagnostic groups vs.\ controls. Positive values indicate increased log-odds of diagnosis relative to controls.}
\label{fig:lasso}
\end{figure}

Figure~\ref{fig:lasso} presents the heatmap of the resulting log-odds ratio estimates, highlighting both shared and diagnosis-specific associations across mood circumplex domains. Anxiousness-related features were consistently associated with all three mood disorders. In particular, both the mean and variability of FPC1 and FPC2 scores showed nonzero associations, suggesting that sustained anxious burden as well as its within-day homeostatic modulation are broadly informative markers of mood pathology. In contrast, energy and TLAC features exhibited disorder-specific patterns. For TLAC, the mean of both FPC1 and FPC2 scores were negatively associated with Bipolar I (FPC1 $\log\text{-OR} = -0.67$, FPC2 $\log\text{-OR} = -0.33$), whereas Bipolar II showed a positive association with the mean of TLAC FPC2 score only ($\log\text{-OR} = 0.15$), indicating differential association of morning to evening homeostatic gradient of activity with the two bipolar subtypes. The variability of TLAC FPC1 scores was found to be positively associated with all the mood disorders.

Energy-related features were selectively associated with MDD. Lower subject-level mean of FPC2  scores ($\log\text{-OR} = -1.12$) and reduced variability of FPC1 ($\log\text{-OR} = -0.82$) were uniquely linked to MDD. These results indicate that individuals with MDD tended to exhibit (i) a more weakly expressed within-day energy gradient and (ii) less day-to-day fluctuation in their overall energy level. Similar attenuations has mainly been documented for \emph{activity}-based rest--activity rhythms in depression \citep{wirzjustice2008diurnal, smagula2015circadian, ho2024actigraphic}. Observing it in {energy} motivates studying activity--energy coupling more directly, since perceived energy plausibly constrains activity (low energy may limit activity even when opportunities exist), pointing to potentially deeper constructs than activity counts alone. Sad mood-based FPC scores did not emerge as a dominant discriminator in the LASSO model, reinforcing the importance of dynamic and temporally structured behavioral signatures from the complete mood circumplex beyond sadness severity for distinguishing mood disorder subtypes.

Together, these results demonstrate the clinical utility of latent functional scores from ps-$M^2$FPCA for phenotyping and risk stratification in mood disorders, a key step toward precision digital psychiatry.

\section{Discussion}\label{sec:discussion}
This paper develops multivariate FPCA for \emph{mixed-type} mHealth trajectories, motivated by joint diurnal patterns in ordinal EMA (sad mood, anxiousness, energy) and continuous actigraphy (TLAC). Unlike the standard multivariate FPCA which typically assumes Gaussian, fully observed continuous functions, our framework accommodates mixed measurement scales and irregular sampling by mapping all modalities to a common latent Gaussian scale via a semiparametric copula. This enables coherent estimation of within and cross-domain temporal dependence and yields interpretable, diagnostically informative digital biomarkers. We posit the Multivariate Generalized Latent Nonparanormal Process (MGLNPP), estimate the latent dependence using Kendall’s $\tau$ bridging, and offer a flexible blockwise covariance estimator ($M^2$FPCA) alongside a scalable partially-separable alternative (ps-$M^2$FPCA). Our empirical analysis using simulations demonstrates the competitive performance of both methods under several complex
dependence structures. 

In the NIMH Family Study, FPC1 captures an overall activation/burden axis linking affective load, energy, and activity, consistent with evidence that physical activity is protective against depression \citep{Schuch2018AJP,Pearce2022JAMAPsych}, while the second component (FPC2) represents a monotonic morning-to-evening contrast consistent with homeostatic sleep pressure. Finally, the third component (FPC3) captures a midday-peaking inverted-U pattern suggestive of circadian modulation of affective and activation states central to bipolar illness \citep{Tonon2024PCN,Scott2022Neubiorev} and measurable through rest-activity rhythms \citep{McGowan2019TranslPsychiatry}.  The joint covariance surfaces and multivariate scores quantify cross-domain coupling in a time-resolved way and complement findings that wearable sleep/circadian features can help predict mood episodes \citep{Lim2024NPJDM}.

Multiple research directions can be explored based on this work. One of the limitations of the current approach is treating subject-days as independent replicates, which facilitates stable extraction of the diurnal components but ignores serial dependence across days.
In future, this motivates an $M^3$FPCA extension for multilevel  mixed-type multivariate functions which will certainly be more computationally challenging. Finally, establishing reliable digital phenotypes will require replication across different devices, study protocols, and populations. Multi-cohort efforts such as the Motor Activity Research Consortium for Health (mMARCH), together with standardized actigraphy pipelines such as GGIR, can support this aim by enabling assessment of whether eigenfunctions, score distributions, and subtype-discriminative biomarker panels are reproducible across studies \citep{Merikangas2025mMARCH}.

\section{Funding}
This study was supported by grant no. ZIA MH002954-07 Motor Activity Research Consortium for Health (mMARCH) from the National Institute of Health. The views and opinions expressed herein are those of the authors and should not be construed to represent the views of any of the sponsoring organizations, agencies, or the US government.


\bibliographystyle{biom} 
\bibliography{ref.bib}

\appendix

\section{Algorithms}
\begin{algorithm}[H]
\caption{$M^2$FPCA via block-wise covariance estimation with weighted Kendall's $\tau$ under irregular sampling}
\label{alg:sg_mfpca_condensed}
\begin{algorithmic}[1]
\Statex \textbf{Input:}
Irregular multivariate functional data $\{(t_{ijr},X_{ij}(t_{ijr})):\ r=1,\ldots,m_{ij}\}$,
basis $\{B_k\}$, bridging functions $\{F_{jk}\}$,
transformation $g(\cdot)$, reliability threshold $c_0$, PD threshold $\epsilon$.
\Statex \textbf{Output:}
Positive definite covariance $\widehat{\mathcal{C}}_{\mathrm{PD}}$,
multivariate eigenfunctions $\{\widehat{\bphi}_\ell\}$,
FPC scores $\{\widehat{\xi}_{i\ell}\}$.

\vspace{1mm}
\Statex \textbf{Stage I: Marginal covariance estimation}
\For{$j=1,\ldots,J$}
    \State Compute pairwise-complete Kendall associations $\widehat{\tau}_{jj}(s,t)$ over $(s,t)\in\calT_{\mathrm{obs}}^2$ and counts $N_{jj}(s,t)$.
    \State Estimate smoothed latent surface $\widehat{C}_{jj}(s,t)$ by weighted least squares using weights $N_{jj}(s,t)$ for $N_{jj}(s,t)>c_0$, with $\widehat{C}_{jj}(t,t)=1$.
    \State Form marginal block $\widehat{\Sigma}_{jj}$.
\EndFor

\vspace{1mm}
\Statex \textbf{Stage II: Cross-covariance estimation}
\For{$1\le j<k\le J$}
    \State Compute pairwise-complete Kendall associations $\widehat{\tau}_{jk}(s,t)$ over $(s,t)\in\calT_{\mathrm{obs}}^2$ and counts $N_{jk}(s,t)$.
    \State Estimate $\widehat{C}_{jk}(s,t)$ by weighted least squares using weights $N_{jk}(s,t)$ for $N_{jk}(s,t)>c_0$.
    \State Form blocks $\widehat{\Sigma}_{jk}$ and $\widehat{\Sigma}_{kj}=\widehat{\Sigma}_{jk}^{\top}$.
\EndFor

\vspace{1mm}
\Statex \textbf{Stage III: Assembly and PD projection}
\State Assemble $\widehat{\mathcal{C}}=(\widehat{\Sigma}_{jk})_{j,k=1}^J$.
\State Eigendecompose $\widehat{\mathcal{C}}=P\Lambda P^{\top}$ and threshold
$\Lambda'_{pp}=\max(\Lambda_{pp},\epsilon)$.
\State Set $\widehat{\mathcal{C}}_{\mathrm{PD}}=P\Lambda'P^{\top}$.

\vspace{1mm}
\Statex \textbf{Stage IV: Latent prediction and multivariate FPCA}
\For{$i=1,\ldots,n$}
    \State Compute latent Gaussian prediction
    $\widehat{\bV}_i=\mathbb{E}(\bV_i\mid\bX_i,\widehat{\mathcal{C}}_{\mathrm{PD}})$.
\EndFor
\State Apply multivariate FPCA to $\{\widehat{\bV}_i\}_{i=1}^n$
using the \texttt{MFPCA} package \citep{MFPCApackage, happ2018multivariate}
to obtain $\{\widehat{\bphi}_\ell,\widehat{\xi}_{i\ell}\}$.
\State \Return $\widehat{\mathcal{C}}_{\mathrm{PD}}$, $\{\widehat{\bphi}_\ell\}$, $\{\widehat{\xi}_{i\ell}\}$.
\end{algorithmic}
\end{algorithm}

\begin{algorithm}[H]
\caption{ps-$M^2$FPCA via $M^1$FPCA latent predictions}
\label{alg:ps_sg_fpca}
\begin{algorithmic}[1]
\Statex \textbf{Input:}
\Statex Mixed-type functional data $\{(t_{ijr},X_{ij}(t_{ijr})):\ i=1,\ldots,n;\ j=1,\ldots,J;\ r=1,\ldots,m_{ij}\}$,
$M^1$FPCA settings (basis, smoothing, truncation levels $M_j$), common grid $S$, ps-$M^2$FPCA truncation level $L$.
\Statex \textbf{Output:}
\Statex Shared eigenfunctions $\{\widehat{\phi}_{\ell}(\cdot)\}_{\ell=1}^{L}$,
score vectors $\{\widehat{\boldsymbol{\bxi}}_{i\ell}\in\mathbb{R}^J\}_{i,\ell}$ (and optional reconstructions).

\vspace{1mm}
\Statex \textbf{Stage I: univariate latent covariance estimation and BLUP prediction}
\For{$j=1$ to $J$}
    \State Estimate the marginal latent correlation/covariance surface $\widehat{C}_{jj}(s,t)$
    via Kendall's $\tau$ bridging and weighted smoothing under irregular sampling (cf.\ \citet{dey2024functional}).
    \State Obtain univariate latent predictions on $S$ by the conditional expectation (univariate BLUP):
    \[
    \widehat{V}_{ij}(t)\leftarrow \mathbb{E}\!\left\{V_{ij}(t)\mid \bX_{ij},\widehat{C}_{jj}\right\},
    \qquad t\in S.
    \]
\EndFor
\State Assemble $\widehat{\bV}_i(t)=(\widehat{V}_{i1}(t),\ldots,\widehat{V}_{iJ}(t))^\top$ for $t\in S$.

\vspace{1mm}
\Statex \textbf{Stage II: partially separable MFPCA on latent predictions}
\State Compute the pooled marginal covariance $\widehat{H}(s,t)\leftarrow J^{-1}\sum_{j=1}^J \widehat{C}_{jj}(s,t)$ on $S\times S$.
\State Obtain shared temporal eigenfunctions $\{\widehat{\phi}_{\ell}\}_{\ell=1}^{L}$ as the leading eigenfunctions of $\widehat{H}$ \citep{zapata2022partial}.
\State For each $\ell=1,\ldots,L$, estimate the score covariance matrix $\widehat{\Sigma}_{\ell}\in\mathbb{R}^{J\times J}$ by projecting the empirical cross-covariances of $\widehat{\bV}_i$ onto $\widehat{\phi}_{\ell}$ (plug-in as in \citet{zapata2022partial}).
\State Compute subject-specific score vectors $\widehat{\boldsymbol{\bxi}}_{i\ell}\in\mathbb{R}^J$ via
\[
\widehat{\boldsymbol{\bxi}}_{i\ell}
\leftarrow
\int_{\calT} \widehat{\bV}_i(t)\,\widehat{\phi}_{\ell}(t)\,dt,
\qquad \ell=1,\ldots,L,
\]
(with numerical integration over $S$).
\end{algorithmic}
\end{algorithm}
\section{Bridging functions for mixed data types}\label{suppl:bridge}  

This appendix lists analytic bridging functions linking population Kendall’s $\tau$ for mixed data types
(continuous, binary, truncated, ordinal) to the latent Gaussian correlation $\rho$ under the semiparametric
Gaussian copula construction. The collection below follows the unified bridging results combining continuous/binary \citep{liu2012high},
truncated \citep{yoon2018sparse}, and ordinal outcomes \citep{dey2022semiparametric}.

\begin{theorem}[Bridging functions for mixed data types]\label{thm:bridging}
Let $X_j$ and $X_k$ be two GLNPP variables generated from an underlying latent bivariate normal vector with
correlation $\rho$. Then the population Kendall’s $\tau$ satisfies $\tau_{jk}=F(\rho)$, where $F$ additionally
depends on the cutoffs $\Delta_j,\Delta_k$ for non-continuous components. The bridging functions corresponding
to all relevant pairs are:
\begin{align}
     F_{\rm cc}(\rho) & = \dfrac{2}{\pi} \sin^{-1}(\rho) \nonumber\\
     F_{\rm bb}(\rho; \Delta_j, \Delta_k)&  = 2\left\{\Phi_2( \Delta_j,  \Delta_{k}; \rho)-\Phi( \Delta_j)\Phi( \Delta_{k})\right\} \nonumber\\
     F_{\rm cb}(\rho; \Delta_j) & = 4\Phi_2( \Delta_j, 0; \rho/\sqrt{2})-2\Phi( \Delta_j) \nonumber\\
     F_{\rm tb}(\rho; \Delta_j, \Delta_k) & =
    2 \Phi_3\left(\Delta_k, -\Delta_j, 0; S_{3a}(\rho) \right)
    -2 \Phi_3\left(\Delta_k, -\Delta_j, 0; S_{3b}(\rho) \right) \nonumber\\
    F_{\rm ct}(\rho; \Delta_j) & = -2 \Phi_2 (-\Delta_j,0; 1/\sqrt{2} ) +4\Phi_3 \left(-\Delta_j,0,0; S_3(\rho)\right) \nonumber\\
    F_{\rm tt}(\rho; \Delta_j, \Delta_k)  & = -2 \Phi_4 (-\Delta_j, -\Delta_k, 0,0; S_{4a}(\rho)) + 2 \Phi_4 (-\Delta_j, -\Delta_k, 0,0; S_{4b}(\rho)) \nonumber\\
    F_{\rm co}(\rho; \Delta_j) & = \sum_{r=1}^{l_j-1} \left\{4 \Phi_3 (\Delta_{jr}, \Delta_{j(r+1)},0; \, S_3(\rho)) - 2 \Phi(\Delta_{jr})\Phi(\Delta_{j(r+1)})\right\} \nonumber\\
    F_{\rm oo}(\rho; \Delta_j, \Delta_k) &= 2 \sum_{r=1}^{l_j-1}\sum_{s=1}^{l_k-1}\Big[\Phi_2(\Delta_{jr},\Delta_{ks}; \rho)\Big\{ \Phi_2(\Delta_{j(r+1)},\Delta_{k(s+1)};\rho) \nonumber\\
&\hspace{35mm} - \Phi_2(\Delta_{j(r+1)},\Delta_{k(s-1)}; \rho)\Big\}\Big]
- 2 \sum_{r=1}^{l_j-1}\Phi({\Delta_{jr}})\Phi_2(\Delta_{j(r+1)},\Delta_{k(l_k-1)}; \rho) \nonumber\\
 F_{\rm ob}(\rho; \Delta_j, \Delta_k) &= 2 \sum_{r=1}^{l_j-1}\Big[\Phi_2(\Delta_{jr},\Delta_{k}; \rho)\Phi(\Delta_{j(r+1)}) - \Phi({\Delta_{jr}})\Phi_2(\Delta_{j(r+1)},\Delta_{k}; \rho)\Big] \nonumber\\
F_{\rm to}(\rho; \Delta_j, \Delta_k) & =2 \Phi_3\big(\Delta_{k(l_k-1)}, -\Delta_j, 0;\, S_{3a}(\rho)\big)
- 2 \sum_{r=1}^{l_{k}-1}\Big[\Phi_4\big(\Delta_{k(r+1)}, \Delta_{kr}, -\Delta_j, 0;\, S_{5}(\rho)\big) \nonumber\\
&\hspace{33mm}- \Phi_4\big(\Delta_{k(r-1)}, \Delta_{kr}, -\Delta_j, 0;\, S_{5}(\rho)\big)\Big].
\label{eqn:bridge-thm1}
\end{align}
Here $\Phi$ denotes the standard normal CDF, and $\Phi_d(\cdot;S)$ denotes the CDF of a $d$-variate standard normal
random vector with correlation matrix $S$. For notational simplicity, we use $\Phi_2(\cdot,\cdot;\rho)$ for the
bivariate normal CDF with correlation $\rho$.
\end{theorem}

\begin{equation}
\begin{split}
S_{3a}(\rho)& =
\begin{pmatrix}
1 & 0 & \rho/\sqrt{2} \\
0 & 1 & 1/\sqrt{2} \\
\rho/\sqrt{2}  & 1/\sqrt{2} & 1
\end{pmatrix}, \qquad
S_{3b}(\rho)=
\begin{pmatrix}
1 & 0 & -1/\sqrt{2} \\
0 & 1 & -\rho/\sqrt{2}\\
-1/\sqrt{2} & -\rho/\sqrt{2} & 1
\end{pmatrix},\\[1mm]
S_3(\rho) & =
\begin{pmatrix}
1 & -\rho & -\rho/\sqrt{2}\\
-\rho & 1 & 1/\sqrt{2}\\
-\rho/\sqrt{2} & 1/\sqrt{2} & 1
\end{pmatrix}, \qquad
S_{4a}(\rho) =
\begin{pmatrix}
1 & 0 & 1/\sqrt{2} & -\rho/\sqrt{2}\\
0& 1 & -\rho/\sqrt{2}& 1/\sqrt{2}\\
1/\sqrt{2}& -\rho/\sqrt{2} & 1& -\rho\\
-\rho/\sqrt{2}& 1/\sqrt{2}& -\rho & 1
\end{pmatrix},\\[1mm]
S_{4b}(\rho) & =
\begin{pmatrix}
1 & \rho & 1/\sqrt{2} & \rho/\sqrt{2}\\
\rho & 1 & \rho/\sqrt{2}& 1/\sqrt{2}\\
1/\sqrt{2}& \rho/\sqrt{2} & 1& \rho \\
\rho/\sqrt{2}& 1/\sqrt{2}& \rho  & 1
\end{pmatrix}, \qquad
S_{5}(\rho)  =
\begin{pmatrix}
1 & 0 &  0 & \rho/\sqrt{2} \\
0 & 1 & -\rho & -\rho/\sqrt{2} \\
0  & -\rho & 1 & 1/\sqrt{2} \\
\rho/\sqrt{2} & -\rho/\sqrt{2} & 1/\sqrt{2} & 1
\end{pmatrix}.
\end{split}
\end{equation}

\section{Supporting tools for Estimation}
\label{suppl:estimation_tools}

\subsection{Bridging functions for Kendall’s $\tau$}\label{sec:bridging_functions}

Estimation of the latent correlation kernel $C$ in \eqref{eq:latent_cov_def} uses the bridging relation - $
\tau_{jk}(s,t)=F_{jk}\{C_{jk}(s,t)\},$, which links population Kendall’s $\tau$ to the latent Gaussian correlation. For any pair $(X_j,X_k)$, we write $\tau_{jk}=F_{jk}(\rho)$, where $\rho$ is the corresponding latent correlation and $F_{jk}$ depends on the marginal data types; when at least one component is binary, ordinal, or truncated, $F_{jk}$ also depends on cutoff parameters. Analytic forms are available for continuous and binary variables \citep{liu2012high}, truncated variables \citep{yoon2018sparse}, and ordinal variables \citep{dey2022semiparametric}. In our setting, we use the appropriate $F_{jk}$ for each component pair; all forms used here are listed in Section \ref{suppl:bridge} of the Supplementary Material.

\subsection{Estimation of cutoff parameters}\label{sec:cutoff_estimation}

Whenever at least one component is binary, ordinal, or truncated, the bridging functions depend on cutoff parameters \citep{liu2012high,dey2022semiparametric}. We estimate the cutoff process $\bDelta(\cdot)$ pointwise from the marginal distributions using the method-of-moments estimator of \citet{dey2022semiparametric}. Because the cutoff process is identifiable only up to a monotone transformation, the estimands are $f_t\{\bDelta(t)\}$; with slight abuse of notation, we continue to denote them by $\bDelta(t)$. For each $t\in\calT$, the estimators are $\widehat{\Delta}(t)=\Phi^{-1}\!\left\{n^{-1}\sum_{i=1}^n I(X_i(t)=0)\right\}$ for binary and truncated components, and $\widehat{\Delta}_k(t)=\Phi^{-1}\!\left\{n^{-1}\sum_{i=1}^n I(X_i(t)\le k-1)\right\}$, $k=1,\ldots,l-1$, for ordinal components.

\subsection{Estimation of monotone transformation functions}\label{sec:transformation_estimation}

The monotone transformations $\{f_t\}$ are not needed for estimating $C$, since Kendall’s $\tau$ is invariant under monotone transformations. They are, however, needed for latent Gaussian prediction for continuous and truncated components and therefore enter both the $M^2$FPCA and ps--$M^2$FPCA procedures. Following \citet{liu2009nonparanormal}, we estimate $f_t$ by $\hat{f}_t(x)=\Phi^{-1}\{\hat{G}_t(x)\}$, where $\hat{G}_t(x)=(n+1)^{-1}\sum_{i=1}^n I\{X_i(t)\le x\}$, with the restriction $x>0$ for truncated variables. For binary and ordinal outcomes, $f_t$ is not separately identifiable and is not estimated \citep{liu2012high,dey2022semiparametric}.
\subsection{Selection of tuning parameters}\label{sec:tuning_parameters}

Implementation of $M^2$FPCA and ps--$M^2$FPCA requires choosing the B-spline basis dimension $K$, the reliability threshold $c_0$, the PD threshold $\epsilon$, and the FPCA truncation levels. We select $K$ using the BIC of \citet{dey2024functional}, $BIC(K)=-2\log\hat L+\{K(K+1)/2\}\log n$, with $-2\log\hat L=n\log|\hat{\bC}|+\sum_{i=1}^n \hat{\*V}_i^T\hat{\bC}^{-1}\hat{\*V}_i+B$, where $\hat{\*V}_i$ is the latent BLUP of $\*V_i$, $\hat{\bC}$ is the estimated correlation matrix, and $B$ is a constant. The threshold $c_0$ excludes time pairs with insufficient pairwise-complete information, $\epsilon$ is fixed at $10^{-3}$ for numerical stability, and the truncation levels ($M_j$ for $M^1$FPCA, $L$ for ps--$M^2$FPCA, and the number of retained multivariate components in $M^2$FPCA) are chosen to explain a prespecified proportion of variation, typically $90\%$--$95\%$.

\section{Additional Figures and Tables}

\begin{figure}[H]
\centering
\includegraphics[width=0.95\textwidth]{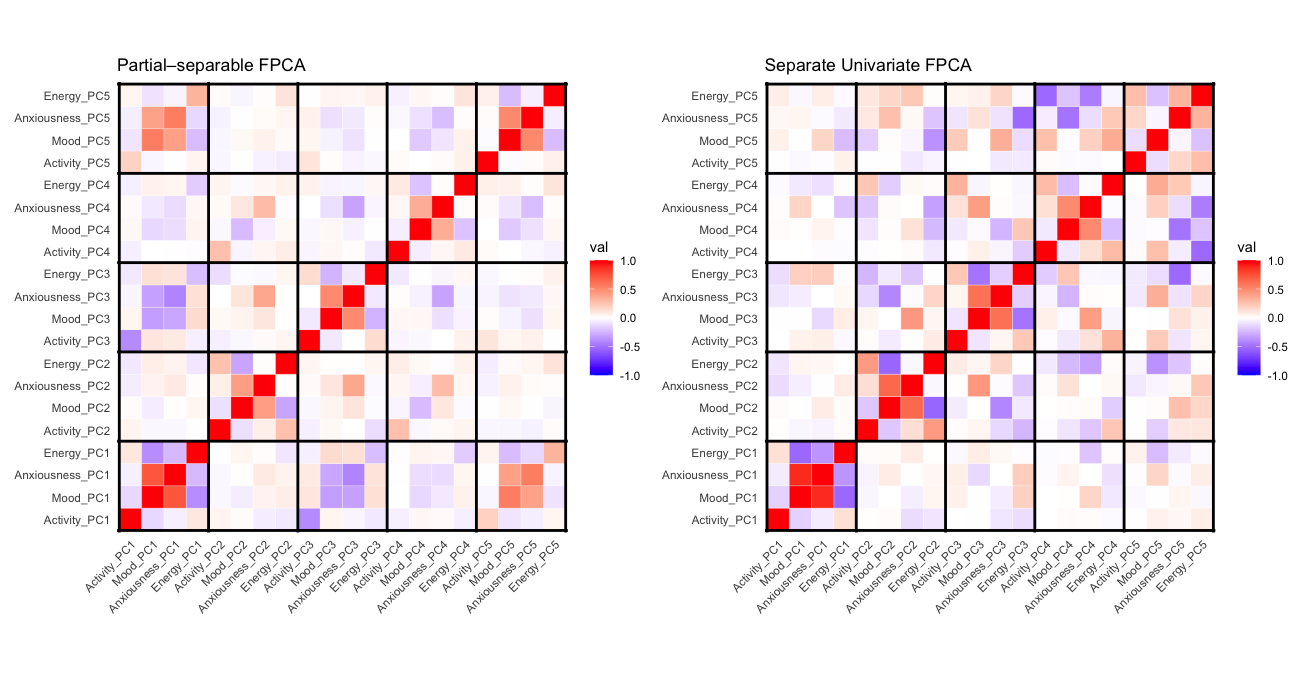}
\caption{Empirical correlation of latent PC scores under partial separability vs.\ univariate FPCA, demonstrating similar dependence structure.}
\label{fig:cov_comp}
\end{figure}

\begin{figure}[H]
\centering
\includegraphics[width=0.65\linewidth , height=0.8\linewidth]{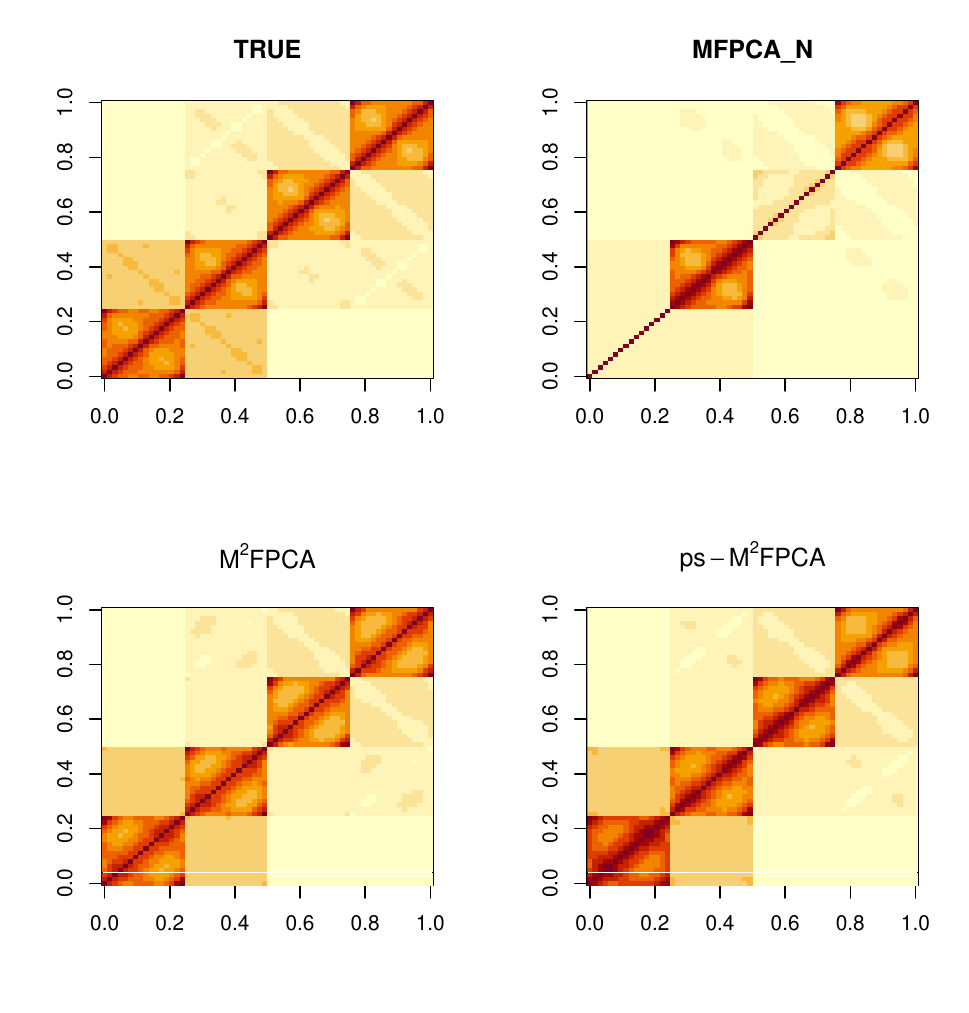}
\caption{True and the Monte-Carlo mean of the estimated covariance surface for the non-stationary correlation kernel and $n=500$, from the naive MFPCA (MFPCA$_N$) and the proposed $M^2$FPCA, ps-$M^2$FPCA.}
\label{fig:fig4new1}
\end{figure}

\begin{figure}[H]
\centering
\includegraphics[width=0.65\linewidth , height=0.8\linewidth]{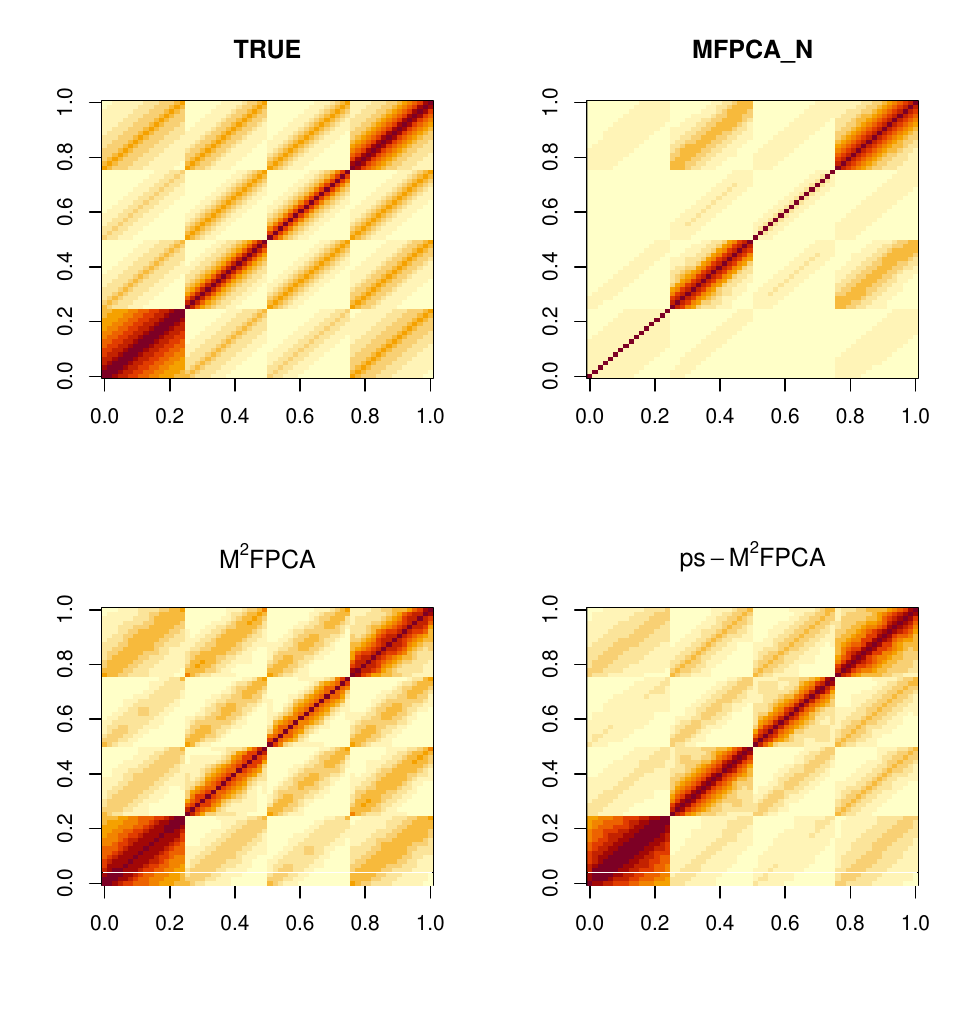}
\caption{True and the Monte-Carlo mean of the estimated covariance surface for the stationary correlation kernel and $n=100$, from the naive MFPCA (MFPCA$_N$) and the proposed $M^2$FPCA, ps-$M^2$FPCA.}
\label{fig:fig5new1}
\end{figure}

\begin{figure}[H]
\centering
\includegraphics[width=0.65\linewidth , height=0.8\linewidth]{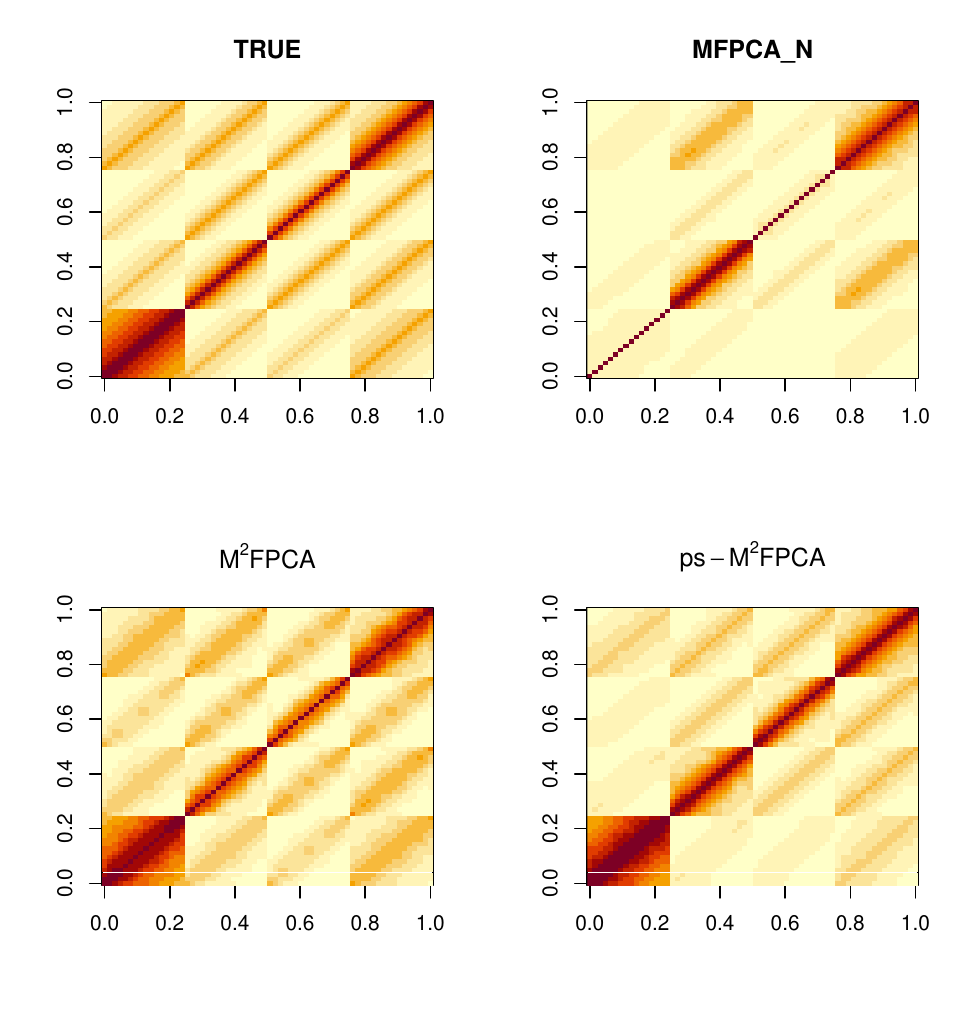}
\caption{True and the Monte-Carlo mean of the estimated covariance surface for the stationary correlation kernel and $n=1000$, from the naive MFPCA (MFPCA$_N$) and the proposed $M^2$FPCA, ps-$M^2$FPCA.}
\label{fig:fig5new2}
\end{figure}

\begin{figure}[H]
\centering
\includegraphics[width=0.65\linewidth , height=0.8\linewidth]{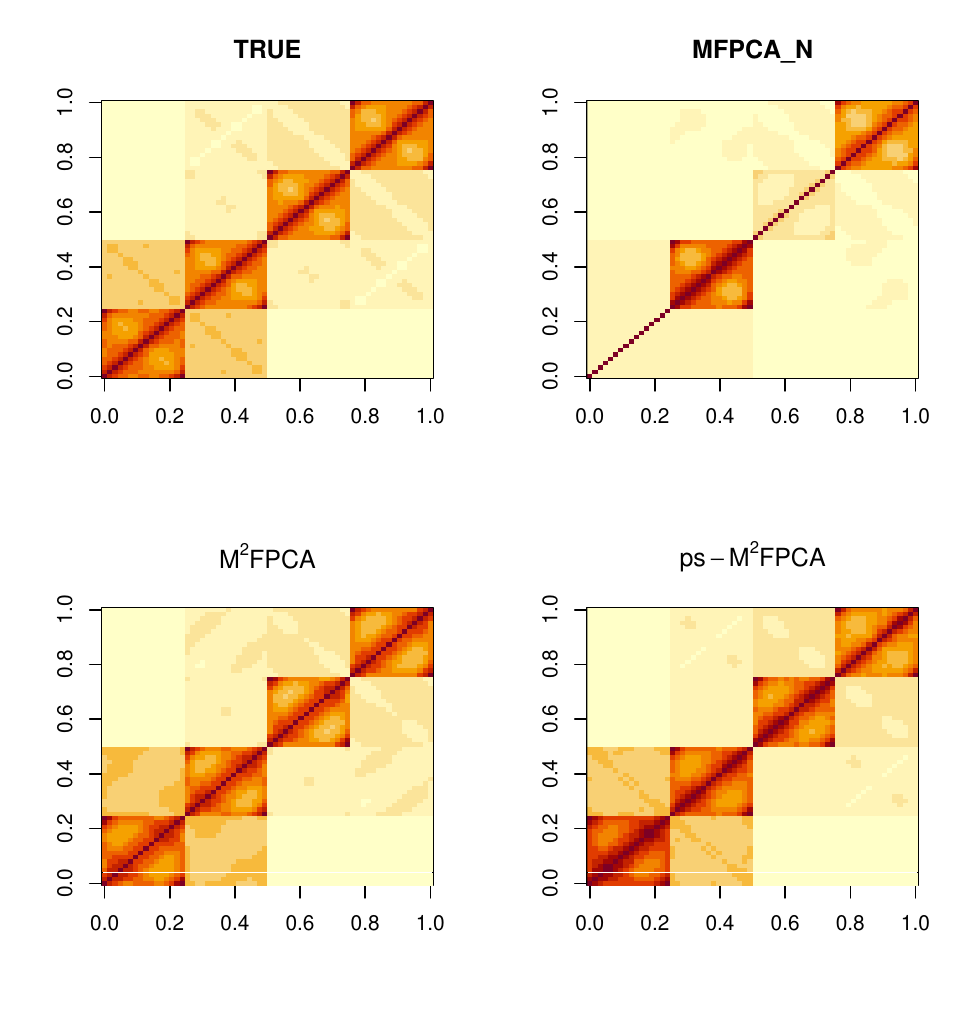}
\caption{True and the Monte-Carlo mean of the estimated covariance surface for the non-stationary correlation kernel and $n=100$, from the naive MFPCA (MFPCA$_N$) and the proposed $M^2$FPCA, ps-$M^2$FPCA.}
\label{fig:fig6new1}
\end{figure}

\begin{figure}[H]
\centering
\includegraphics[width=0.65\linewidth , height=0.8\linewidth]{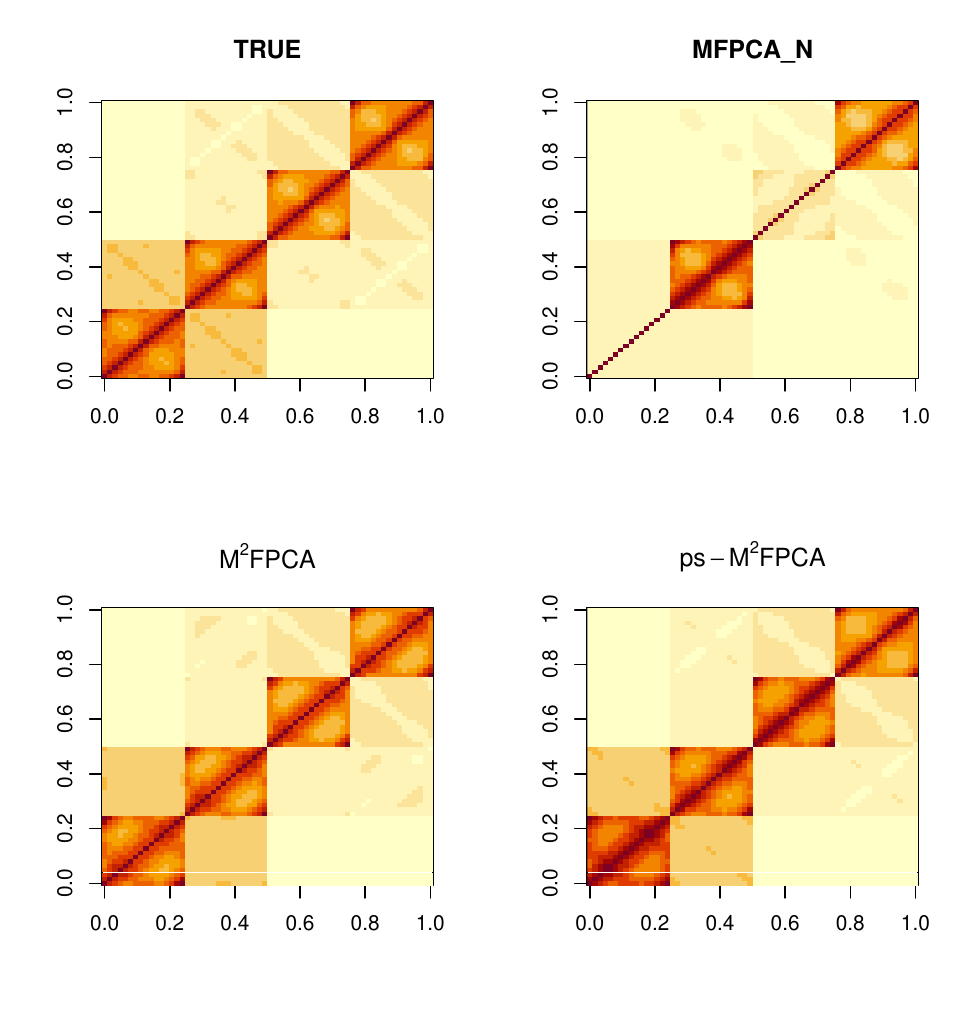}
\caption{True and the Monte-Carlo mean of the estimated covariance surface for the non-stationary correlation kernel and $n=1000$, from the naive MFPCA (MFPCA$_N$) and the proposed $M^2$FPCA, ps-$M^2$FPCA.}
\label{fig:fig6new2}
\end{figure}

\begin{figure}[!t]
\centering
\includegraphics[width=0.95\textwidth]{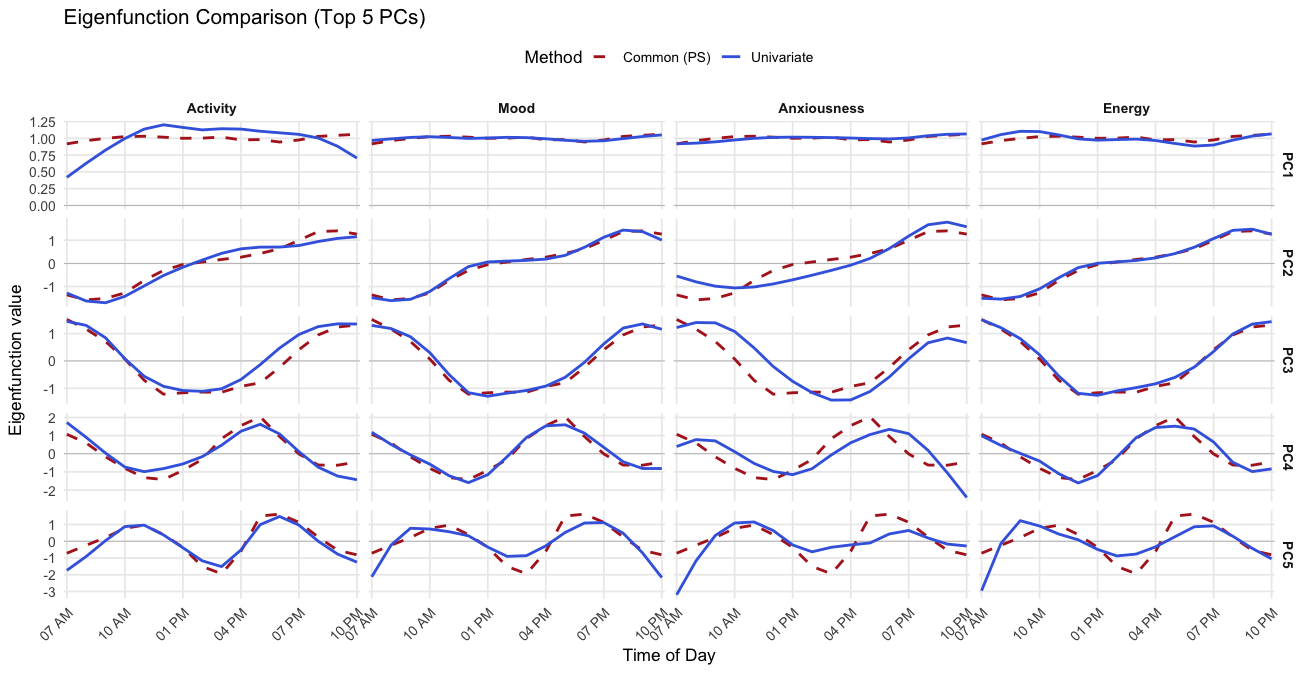}
\caption{Comparison of partial separable vs.\ univariate eigenfunctions across four domains (Activity, Sad Mood, Anxiousness, Energy).}
\label{fig:eigen_comp}
\end{figure}

\begin{table}[H]
\centering
\caption{Baseline characteristics and EMA summaries by diagnostic group.}
\label{tab:table1_baseline}
\fontsize{8pt}{10pt}\selectfont

\begin{adjustbox}{width=\textwidth}
\begin{tabular*}{\textwidth}{@{\extracolsep{\fill}}lccccc c}
\toprule
 & \multicolumn{5}{c}{\textbf{Diagnosis group}} & \\ 
\cmidrule(lr){2-6}
\textbf{Characteristic} 
& \textbf{Overall} 
& \textbf{Control} 
& \textbf{Bipolar I} 
& \textbf{Bipolar II} 
& \textbf{MDD} 
& \textbf{p-value}\textsuperscript{\textit{2}} \\

& N = 307\textsuperscript{\textit{1}}
& N = 130\textsuperscript{\textit{1}}
& N = 41\textsuperscript{\textit{1}}
& N = 30\textsuperscript{\textit{1}}
& N = 106\textsuperscript{\textit{1}}
& \\ 
\midrule

\textbf{Age} 
& 43.85 (18.97) 
& 43.40 (21.22) 
& 43.83 (12.53) 
& 35.77 (17.98) 
& 46.70 (17.88) 
& 0.059 \\ 

\textbf{Sex} 
&  
&  
&  
&  
&  
& 0.035 \\ 
\quad Male   
& 115 (37\%) 
& 61 (47\%) 
& 12 (29\%) 
& 10 (33\%) 
& 32 (30\%) 
&  \\ 
\quad Female 
& 192 (63\%) 
& 69 (53\%) 
& 29 (71\%) 
& 20 (67\%) 
& 74 (70\%) 
&  \\ 

\textbf{Sad mood} 
& 2.45 (0.87) 
& 2.22 (0.82) 
& 2.73 (0.91) 
& 2.73 (0.88) 
& 2.55 (0.86) 
& $<0.001$ \\ 

\textbf{Anxiousness} 
& 2.48 (0.95) 
& 2.23 (0.88) 
& 2.74 (0.93) 
& 2.93 (1.01) 
& 2.56 (0.93) 
& $<0.001$ \\ 

\textbf{Energy} 
& 3.75 (0.86) 
& 3.95 (0.84) 
& 3.52 (0.82) 
& 3.57 (0.85) 
& 3.65 (0.86) 
& $<0.001$ \\ 

\textbf{Sad mood (avg)} 
& 1.73 (0.89) 
& 1.48 (0.84) 
& 2.04 (0.96) 
& 2.02 (0.87) 
& 1.84 (0.86) 
& $<0.001$ \\ 

\textbf{TLAC} 
& 3{,}961.39 (673.26) 
& 4{,}014.92 (689.95) 
& 3{,}646.37 (654.50) 
& 3{,}892.38 (656.09) 
& 4{,}037.13 (636.09) 
& 0.004 \\ 

\textbf{Days per person} 
& 11.31 (2.39) 
& 11.30 (2.40) 
& 11.17 (2.60) 
& 10.87 (2.98) 
& 11.51 (2.12) 
& 0.8 \\ 

\bottomrule
\end{tabular*}
\end{adjustbox}

\vspace{0.4em}
\begin{minipage}{0.95\textwidth}
\footnotesize
\textsuperscript{\textit{1}}Values are mean (SD) for continuous variables and $n$ (\%) for categorical variables.\\
\textsuperscript{\textit{2}}P-values are from Kruskal--Wallis tests for continuous variables and Fisher’s exact tests for categorical variables.
\end{minipage}
\end{table}

\end{document}